\documentclass[useAMS,usenatbib]{mnras}

\usepackage{graphicx}
\usepackage{subfig}
\usepackage{natbib}
\usepackage{ulem}
\usepackage{subfig}
\usepackage{float}
\usepackage{multirow,multicol}
\usepackage{footnote}
\usepackage[bottom]{footmisc}
\usepackage{ulem}
\usepackage{xltabular}
\usepackage{longtable}
\usepackage{lipsum} 
\usepackage{pdflscape}
\usepackage{hyperref}
\usepackage{xcolor, colortbl}
\usepackage{lastpage}
\usepackage{verbatim}
\hypersetup{
colorlinks = true, 
urlcolor = magenta, 
linkcolor = blue, 
citecolor = blue 
}
\usepackage[all]{hypcap}
\usepackage{txfonts}

\newcommand{\teff}{T_\mathrm{eff}}
\newcommand{\logg}{\mathrm{log}\,g}
\newcommand{\iraf}{\textsc{iraf}}

\usepackage{numdef}
\num\def{\lp8}{LP~877-72}
\num\def{\lp1}{LP~1033-31}
\definecolor{bostonuniversityred}{rgb}{0.8, 0.0, 0.0}
\def\deg{\hbox{$^\circ$}} 
\def\lsun{$\rm{L}_{\odot}$} 
\def\msun{$\rm{M}_{\odot}$} 
\def\rsun{$\rm{R}_{\odot}$} 
\newcommand{\pcite}[1]{\protect\cite{#1}} 

\newcommand{\gsimeq}{\hbox{\raise0.5ex\hbox{$>\lower1.06ex\hbox{$\kern-0.92em{\sim}$}$}}}
\newcommand{\lsimeq}{\hbox{\raise0.5ex\hbox{$<\lower1.06ex\hbox{$\kern-0.92em{\sim}$}$}}}
\title[Detection and characterisation of two VLM binaries]{Detection and characterisation of two VLM binaries: \lp1\ and \lp8\thanks{Based on observations made with the NaCo instrument on the VLT@ESO telescope at paranal Observatory under programme ID 091.C-0501(B).}}

\author[Karmakar et al.]{Subhajeet Karmakar$^{1}$\thanks{E-mail: \href{mailto:subhajeet09@gmail.com}{subhajeet09@gmail.com}, subhajeet@prl.res.in}, A. S. Rajpurohit$^{1}$, F. Allard$^{2}$, and D. Homeier$^{3}$\\
$^{1}$ Astronomy \& Astrophysics Division, Physical Research Laboratory, Navrangapura, Ahmedabad 380009, India\\
$^{2}$ Univ Lyon, Ens de Lyon, Univ Lyon1, CNRS, Centre de Recherche Astrophysique de Lyon UMR5574, F-69007, Lyon, France\\
$^{3}$ Zentrum f\"{u}r Astronomie der Universit\"{a}t Heidelberg, Landessternwarte, K\"{o}nigstuhl 12, 69117 Heidelberg, Germany\\
}

\begin{document} 
\pagerange{\pageref{firstpage}--\pageref{LastPage}} \pubyear{2020}
\maketitle
\label{firstpage}

\begin{abstract}
Using the high-resolution near-infrared adaptive optics imaging from the NaCo instrument at the Very Large Telescope, we report the discovery of a new binary companion to the M-dwarf \lp1\ and also confirm the binarity of \lp8. We have characterised both the stellar systems and estimated the properties of their individual components. We have found that \lp1~AB with the spectral type of M4.5+M4.5 has a projected separation of 6.7$\pm$1.3~AU. Whereas with the spectral type of M1+M4, the projected separation of \lp8~AB is estimated to be 45.8$\pm$0.3~AU. The binary companions of \lp1~AB are found to have similar masses, radii, effective temperatures, and log~$g$ with the estimated values of 0.20$\pm$0.04~\msun, 0.22$\pm$0.03~\rsun, 3200 K, 5.06$\pm$0.04. However, the primary of \lp8~AB is found to be twice as massive as the secondary with the derived mass of 0.520$\pm$0.006~\msun. The radius and log~$g$ for the primary of \lp8~AB are found to be 1.8 and 0.95 times that of the secondary component with the estimated values of 0.492$\pm$0.011~\rsun\ and 4.768$\pm$0.005, respectively. With an effective temperature of  3750$\pm$15~K, the primary of \lp8~AB is also estimated to be $\sim$400~K hotter than the secondary component. We have also estimated the orbital period of \lp1\ and \lp8\ to be $\sim$28 and $\sim$349~yr, respectively. The binding energies for both systems are found to be $>$10$^{43}$~erg, which signifies both systems are stable. 
\end{abstract}

\section{Introduction}
\label{sec:intro}
Most of the stars in the galactic stellar population consist of low mass stars \citep{Bochanski2010}. These stars show high a level of magnetic activities due to their thick convective envelope above a radiative interior \citep[][]{Pandey-15-AJ-6, Savanov-16-AcA-3, Karmakar-16-MNRAS-8, Karmakar-17-ApJ-5}. Among these stars, M-dwarfs are the most abundant inhabitants of our Galaxy. These stars also account for over 70\% of stellar systems in the solar neighbourhood \citep[][]{Henry.-97-AJ-9}. Further, M-dwarfs are the most numerous potential planet hosts of all the stellar classes \citep[][]{LadaC-06-ApJ-3}. The typical mass of the very low mass (VLM) stars or M-dwarfs ranges from 0.6 $M_{\odot}$ to the hydrogen-burning limit of about 0.075 $M_{\odot}$ \citep[][]{BaraffeI-98-A+A-1}. 
In this paper, the `VLM binaries' are used to denote the binary systems that consist of VLM+VLM stellar companions.

From the past few decades, following the first discoveries by \cite{Nakajima.-95-Natur-3} and \cite{Rebolo.-95-Natur-1}, thousands of VLM stars have been discovered till date. These stars are intrinsically faint due to their small size and low temperature. However, with the advent of large infrared arrays, the number of VLM binaries that have been detected has increased substantially in recent years. Various ground-based surveys such as Sloan Digital Sky Survey \citep[SDSS;][]{York2000}, Two Micron All Sky Survey \citep[2MASS;][]{Cutri-03-yCat-4, Skrutskie2006}, and space-based surveys such as Wide-field Infrared Survey Explorer \citep[WISE;][]{Wright2010} have made such discoveries. However, despite the large assemblage of these cool objects, their formation mechanism still remains an open question \citep[e.g.][]{BurgasserA-07-ApJ-5, LuhmanK-07-prpl-8, WhitworthA-07-prpl-2, LuhmanK-12-ARA+A-5, ChabrierG-14-prpl-2}. In order to explain their origins, various scenarios such as ejection from multiple prestellar cores \citep[e.g.][]{Reipurth-01-AJ-4}, turbulent fragmentation of gas in protostellar clouds \citep[e.g.][]{PadoanP-04-ApJ-1}, photoionizing radiation from massive nearby stars \citep[e.g.][]{WhitworthA-04-A+A}, and fragmentation of unstable prestellar disks \citep[e.g.][]{StamatellosD-09-MNRAS-4} have been proposed. 

The observational studies of VLM binaries can provide effective diagnostics for testing the VLM formation scenarios. This is due to the fact that the formation mechanisms leave their own traces on the statistical properties of binaries such as frequency, orbit separation, and mass-ratio distributions \citep[e.g.][]{BateM-09-MNRAS-1}. In addition, VLM binaries can provide a model-independent way to determine physical properties, including masses and radii \citep[see][]{Delfosse-00-A+A-3, Lane-01-ApJ-5, Segransan2003, BouyH-04-A+A-3, KonopackyQ-10-ApJ-1, DupuyT-17-ApJS-4, Winters-19-AJ-1, Winters-19-AJ-4}. This is fundamental to the calibration of the mass-luminosity relation \citep[][]{HenryT-93-AJ-2, HenryT-99-ApJ-4, SegransanD-00-A+A-1}. Hence, for the comprehensive study of the formation scenarios, it is essential to obtain unbiased VLM binary samples from various detection methods that are effective with respect to a different population of VLM objects. 

The small size of VLMs makes them suitable candidates to detect planets around them in the habitable zone. Recent studies by \cite{Bonfils-12-A+A-1}, \cite{Anglada2016}, and \cite{Gillon2017} show that M-dwarfs host exoplanets. Recently, few large exoplanet surveys have been started to monitor sizeable numbers of M-dwarfs, such as the M2K programme which is targeting some 1600 M-dwarfs for radial velocity (RV) monitoring \citep[][]{AppsK-10-PASP}, the CARMENES search for exoplanets around 324 M-dwarfs \citep[][]{ReinersA-18-A+A-2}, and the MEarth project \citep[][]{IrwinJ-15-csss-7} which is designed to detect exoplanet transits in nearby late-type M-dwarfs. Among these studies, the brightest M-dwarfs are the ideal (highest priority) targets for high-precision RV searches, whereas the latest M-dwarfs are the most suitable for transit surveys, and the youngest M-dwarfs are preferable targets for adaptive optics (AO) imaging.
In the case of VLM binaries, the presence of the companion to the primary star is also very important as they influence planet formation. However, due to limited detection of exoplanetary systems in the binary or multiple systems particularly on VLM stellar systems, the studies of the effect of stellar multiplicity on the planet formation is still statistically insignificant. For example, in case of the stellar systems in the solar neighbourhood, \cite{WangJ-14-ApJ-24} have shown that compared to the single-star systems, planets in multiple-star systems occur 4.5$\pm$3.2, 2.6$\pm$1.0, and 1.7$\pm$0.5 times less frequently when a stellar companion is present at a distance of 10, 100, and 1000 AU, respectively. Therefore, further observations of the VLM binaries are essential.

The high spatial resolution imaging surveys in recent years enables us to estimate the binary separation for VLM stars. Several studies regarding the binary separation in the last few decades show significant progress. \cite{Phan-Bao-05-A+A-3} have summarised the binary frequency in the separation range 1--15 AU is about 15\%, whereas the frequency of wide binary systems (semi-major axis $>$15 AU) is very low, $<$1\%. Moreover, the mass ratios were strongly biased towards nearly-equal mass binaries. Other studies showed that from a statistical sample with separation over 3 AU, the typical binary separation was found to be $\sim$4 AU \citep[e.g.][]{Close-03-ApJ-4, BurgasserA-07-prpl-7, KrausA-12-ApJ-7}. Via statistical investigation utilising a Bayesian algorithm, \cite{AllenP-12-AJ} found that only 2.3 \% of VLM objects have a companion in the 40--1000 AU range. Also, \cite{BaronF-15-ApJ-2} reported the discovery of 14 VLM binary systems with separations of 250--7500 AU. The studies of \cite{Galvez-Ortiz-17-MNRAS} reported the identification of 36 low and VLM candidates to binary/multiple systems with separations between 200 and 92\,000 AU. Recently, \cite{Winters-19-AJ-4} have derived the projected linear separation distribution of the companions for lower mass primaries to be in the range of 4--20 AU. These differences can be due either to a continuous formation mass-dependent trend or to differences in the formation mechanism of the VLM objects.

Since the proper motion (PM) of a star is inversely proportional to its distance from the observer, a high proper motion (HPM) is a strong selection criterion for proximity.
The first attempts at large surveys for HPM stars began in the early 20th century with works by \cite{van-MaanenA-15-PASP-1}, \cite{WolfM-19-VeHei-20}, and \cite{RossF-39-AJ-1}. Later, additional surveys were completed: e.g. \cite{GiclasH-71-lpms-1, GiclasH-78-LowOB-6}. The first all-sky search for nearby stars was carried out by Luyten, who published two PM catalogues: the Luyten Half-Second catalogue \citep[][]{LuytenW-79-lccs-7} and the New Luyten Two-Tenths catalogue \citep[NLTT;][]{LuytenW-79-nlcs-5, LuytenW-80-nltt-5}. Luyten's catalogues represent the results of a huge effort over more than three decades and are mainly based on observations with the Palomar Schmidt telescope and the blinking technique applied to pairs of plates with approximately 15 years epoch difference. 
After these early works, many studies on PM have been carried out with different galactic latitude, source magnitude, and proper motion range \citep[][]{LepineS-05-AJ-7, LepineS-08-AJ-4, BoydM-11-AJ-3, LepineS-11-AJ-3, LepineS-13-AJ-1, LepineS-13-AN-2, FrithJ-13-MNRAS-1, SmithL-14-MNRAS-9, LuhmanK-14-ApJ-9, KirkpatrickJ-14-ApJ-3, KurtevR-17-MNRAS-2}. 
\begin{figure*}
\centering
\subfloat[LP~1033-31]{\includegraphics[height=8.1cm, angle=0]{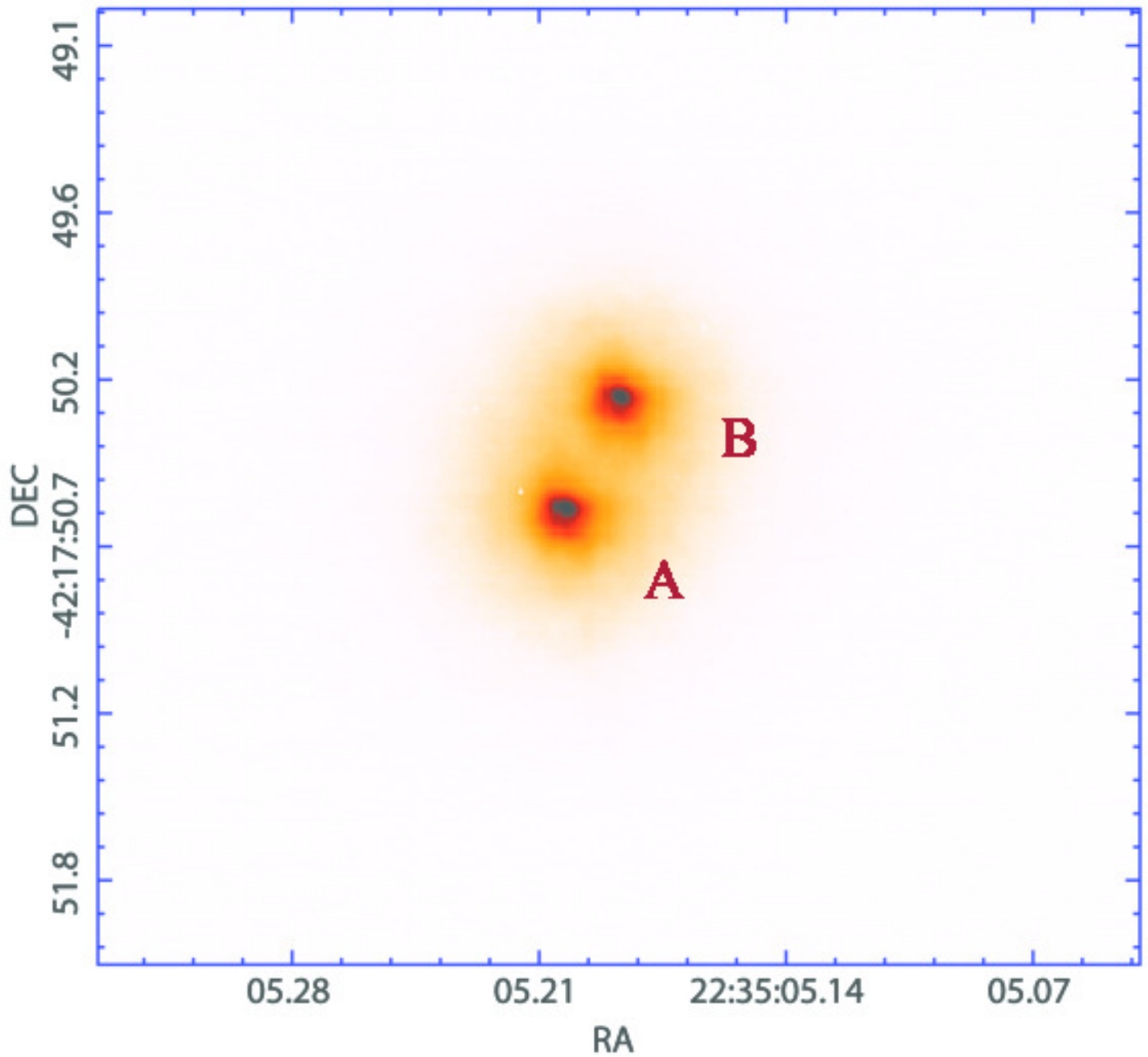}}
\hspace{0.2mm}
\subfloat[LP~877-72]{\includegraphics[height=8cm, angle=0]{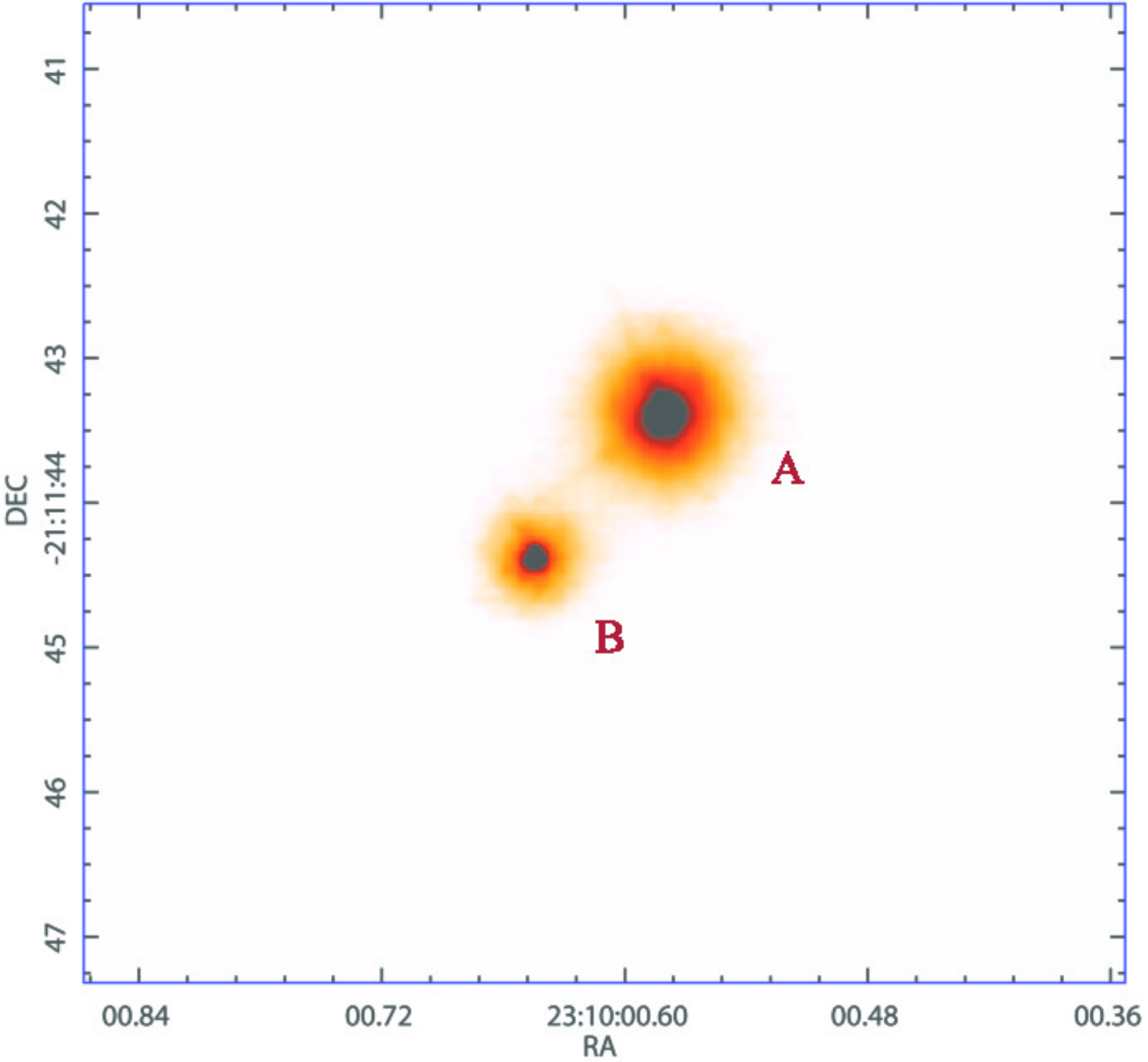}}

\caption{NaCo $J$-band images of (a) \lp1\ and (b) \lp8. The standard convention is used, i.e. North is up, and East is left. A and B denotes the primary and secondary components of both the VLM binaries.}
\label{fig:Image}
\end{figure*}

In this paper, we report near-infrared (NIR) detection of VLM companions to two M-dwarfs \lp1\ and \lp8\ using direct imaging and characterised them. We have discovered that \lp1\ is a binary system instead of a `single' object as known previously. We have also confirmed the binarity of \lp8\ after the \textit{Gaia} detection of the companion \citep[][]{Gaia-Collaboration-18-A+A-5}. Both the objects are far from the galactic plane, and they were first detected as part of HPM surveys \citep[][]{GiclasH-78-LowOB-6, LuytenW-95-yCat-1}. Later these objects were also observed as the part of several high proper-motion surveys \citep[][]{WroblewskiH-95-A+AS, PokornyR-03-A+A, SalimS-03-ApJ-3, ReyleC-04-A+A-2, SchneiderA-16-ApJ-8, KirkpatrickJ-16-ApJS-3} The latest estimated values for proper motion of the systems are 0\arcsec $\!$.299 and 0\arcsec $\!$.305 per year for \lp1\ and \lp8, respectively \citep[][]{Muirhead.-18-AJ-3, StassunK-19-AJ-2}. Both of these objects were also studied as a part of the bright M-dwarfs \citep[][]{LepineS-11-AJ-3, FrithJ-13-MNRAS-1}, as well as the solar-neighbourhoods \citep[][]{Finch-14-AJ-8, WintersJ-15-AJ-2}.

Several investigations of the stellar parameters of \lp1\ and \lp8\ have been performed over the last few decades. The first spectroscopic observations for \lp1\ and \lp8\ as `single' objects have been performed by \cite{Reyle-06-MNRAS-3} and \cite{Scholz-05-A+A-9}, respectively. Using visual comparison with the spectral templates as well as from the classification scheme based on the TiO and CaH band strengths, \cite{Reyle-06-MNRAS-3} derived the spectral for \lp1\ to be M3.5. Using the same method, \cite{Scholz-05-A+A-9} derived the spectral type for \lp8\ to be M3.0. 
\cite{Rajpurohit-13-A+A-2} had performed the spectral synthesis analysis to determine their atmospheric properties and their fundamental parameters such as $\teff$ and log~$g$ by assuming the solar metallicity. The derived $\teff$, spectral type, and log~$g$ were found to be equal for both \lp1\ and \lp8\ with the values of 3200 K, M4, and 5.0. Later observations of these parameters show good agreement with these values \citep[][]{StassunK-18-AJ-2, StassunK-19-AJ-2, Anders.-19-A+A-2}. 
Using TESS data, as `single' object the mass and radius of \lp1\ have been derived as 0.316 \msun\ and 0.319 \rsun\ \citep[][]{StassunK-19-AJ-2}, whereas for \lp8\ these same parameters have been derived as 0.651 \msun\ and 0.556 \rsun\ \citep[][]{Anders.-19-A+A-2, StassunK-19-AJ-2}. Moreover, in the case of \lp8, \cite{Gaidos.-14-MNRAS-1} have discussed the possibility of hosting exoplanets and life. 
In this work, we have adopted the distance of \lp8\ derived from the parallax measurement from \textit{Gaia} DR2 to be 33.77$\pm$0.17 pc \citep[][]{Bailer-Jones-18-AJ-6}. However, in the case of \lp1, the \textit{Gaia} parallax measurement is not available. Therefore, we have used the latest photometric distance of \lp1\ as 16.57$\pm$3.27 pc \citep[][]{Finch-14-AJ-8}.
Although \textit{Gaia} observations have already detected both the components of \lp8, the detailed study of both the components was not previously carried out. Therefore, in this study, we discovered the companion of \lp1\ using the high-resolution images. Further, we have characterised both the systems and derived the physical parameters of the identified components.

The paper is organised as follows. The observations are described in Section~\ref{sec:obs}. In Section~\ref{sec:datared}, we discuss the data reduction procedures and ascertain the binary separation, flux-ratio, and position angles. In Section~\ref{sec:nir_photometry} and \ref{sec:sptype}, we have derived the NIR magnitudes and spectral classes of the newly discovered stellar components. In Section~\ref{sec:parameters}, we describe the other derived system-parameters such as mass, radius, and orbital period. In Section~\ref{sec:discussion}, we discuss our results and finally in Section~\ref{sec:summary}, we summarised our work.

\section{Observations}
\label{sec:obs}
Using the Very Large Telescope (VLT) of European Southern Observatory (ESO, Chile) specifically the Nasmyth Adaptive Optics System (NAOS) and Near-Infrared Imager and Spectrograph (CONICA), popularly known as NaCo \citep[][]{Lenzen-03-SPIE, Rousset-03-SPIE}, two M-dwarfs \lp1\ and \lp8\ were observed under programme ID 091.C-0501(B). 
A total of 20 frames were observed in each of the $JHK_s$ photometric bands for \lp1\ on 2013 July 02. On the same night, \lp8\ was also observed for 30 frames in each of the $JHK_s$ photometric bands. The exposure time for individual frames was of 30~s. 
In order to allow an efficient reduction of the sky background, each image was jittered using a jitter box of 5\arcsec. 
Since no bright adaptive optics (AO) source was available nearby, we used N90C10 dichroic where 90\% of the light was used for AO and only 10\% for the science camera. 
All the observations for \lp1\ were obtained within an airmass range of 1.049 -- 1.051, whereas \lp8\ were observed within the airmass ranging from 1.002 -- 1.010. The seeing condition for the day of observation was also very good ($<$ 0.85\arcsec). S13 camera was used with a field of view of $14\arcsec \times 14\arcsec$ and a pixel scale of 13.2~mas per pixel.
\begin{table*}
\setlength{\tabcolsep}{1.6mm}
\center
\caption{\label{tab:measurements} Binary separation, position angle, and flux-ratio measurements for the observed 60 images for \lp1\ and 90 images for \lp8.}
\begin{tabular}{cccc|cccc|cccc}
\hline
\hline
\noalign{\smallskip}
\multicolumn{4}{c|}{$J$-Band} &\multicolumn{4}{c|}{$H$-Band} &\multicolumn{4}{c}{$K_s$-Band} \\
\noalign{\smallskip}
\hline
\noalign{\smallskip}
\# & Separation & PA & Flux ratio & \# & Separation & PA & Flux ratio & \# & Separation & PA & Flux ratio \\
& [mas] &&& & [mas] &&& & [mas] &&\\
\noalign{\smallskip}
\hline
\noalign{\smallskip}
\noalign{\smallskip}
\multicolumn{12}{c}{\textbf{\lp1}}\\
\noalign{\smallskip}
\hline
\hline
\noalign{\smallskip}

1 & 398.4 & 334.08\deg & 1.06 $\pm$ 0.03 & 1 & 402.7 & 333.94\deg & 1.07 $\pm$ 0.04 & 1 & 400.5 & 334.04\deg &1.08 $\pm$ 0.04 \\
2 & 400.8 & 333.83\deg & 1.06 $\pm$ 0.07 & 2 & 399.9 & 333.76\deg & 1.07 $\pm$ 0.02 & 2 & 402.0 & 334.14\deg &1.07 $\pm$ 0.03 \\
3 & 402.3 & 334.14\deg & 1.07 $\pm$ 0.03 & 3 & 399.6 & 333.94\deg & 1.07 $\pm$ 0.09 & 3 & 400.0 & 333.99\deg &1.05 $\pm$ 0.09 \\
4 & 401.6 & 333.91\deg & 1.06 $\pm$ 0.07 & 4 & 401.7 & 334.21\deg & 1.07 $\pm$ 0.05 & 4 & 400.3 & 334.09\deg &1.07 $\pm$ 0.08 \\
5 & 402.4 & 334.46\deg & 1.06 $\pm$ 0.03 & 5 & 402.7 & 333.90\deg & 1.07 $\pm$ 0.03 & 5 & 399.0 & 334.17\deg &1.05 $\pm$ 0.07 \\
6 & 401.1 & 334.13\deg & 1.06 $\pm$ 0.08 & 6 & 400.8 & 334.09\deg & 1.06 $\pm$ 0.03 & 6 & 402.4 & 334.07\deg &1.06 $\pm$ 0.06 \\
7 & 402.2 & 334.12\deg & 1.07 $\pm$ 0.04 & 7 & 399.7 & 333.99\deg & 1.07 $\pm$ 0.04 & 7 & 400.8 & 334.06\deg &1.06 $\pm$ 0.06 \\
8 & 403.6 & 333.96\deg & 1.05 $\pm$ 0.04 & 8 & 400.8 & 333.96\deg & 1.07 $\pm$ 0.04 & 8 & 400.2 & 333.96\deg &1.07 $\pm$ 0.05 \\
9 & 403.4 & 333.97\deg & 1.08 $\pm$ 0.03 & 9 & 399.8 & 333.91\deg & 1.06 $\pm$ 0.08 & 9 & 402.5 & 334.08\deg &1.06 $\pm$ 0.08 \\
10 & 403.3 & 334.02\deg & 1.07 $\pm$ 0.03 & 10 & 400.0 & 334.04\deg & 1.06 $\pm$ 0.04 & 10 & 402.4 & 333.92\deg &1.07 $\pm$ 0.07 \\
11 & 404.8 & 334.13\deg & 1.07 $\pm$ 0.04 & 11 & 400.2 & 334.04\deg & 1.08 $\pm$ 0.05 & 11 & 402.0 & 334.04\deg &1.06 $\pm$ 0.04 \\
12 & 400.8 & 333.92\deg & 1.06 $\pm$ 0.06 & 12 & 403.4 & 333.80\deg & 1.06 $\pm$ 0.03 & 12 & 401.1 & 334.09\deg &1.06 $\pm$ 0.08 \\
13 & 403.8 & 333.91\deg & 1.07 $\pm$ 0.04 & 13 & 400.6 & 333.84\deg & 1.06 $\pm$ 0.09 & 13 & 400.9 & 334.30\deg &1.05 $\pm$ 0.13 \\
14 & 400.5 & 333.98\deg & 1.07 $\pm$ 0.04 & 14 & 401.9 & 334.03\deg & 1.06 $\pm$ 0.04 & 14 & 402.1 & 333.88\deg &1.05 $\pm$ 0.04 \\
15 & 403.1 & 333.93\deg & 1.06 $\pm$ 0.04 & 15 & 400.1 & 334.01\deg & 1.07 $\pm$ 0.04 & 15 & 402.0 & 334.13\deg &1.06 $\pm$ 0.06 \\
16 & 403.8 & 333.79\deg & 1.06 $\pm$ 0.03 & 16 & 402.1 & 333.92\deg & 1.06 $\pm$ 0.03 & 16 & 402.7 & 334.11\deg &1.07 $\pm$ 0.04 \\
17 & 402.9 & 333.66\deg & 1.07 $\pm$ 0.03 & 17 & 401.5 & 333.86\deg & 1.06 $\pm$ 0.04 & 17 & 401.8 & 333.94\deg &1.07 $\pm$ 0.03 \\
18 & 401.2 & 334.07\deg & 1.08 $\pm$ 0.08 & 18 & 402.9 & 333.91\deg & 1.07 $\pm$ 0.03 & 18 & 400.3 & 334.21\deg &1.06 $\pm$ 0.11 \\
19 & 402.6 & 333.91\deg & 1.05 $\pm$ 0.04 & 19 & 400.7 & 333.97\deg & 1.07 $\pm$ 0.04 & 19 & 400.1 & 334.18\deg &1.05 $\pm$ 0.08 \\
20 & 400.6 & 333.86\deg & 1.06 $\pm$ 0.03 & 20 & 401.7 & 333.89\deg & 1.06 $\pm$ 0.06 & 20 & 401.4 & 334.06\deg &1.06 $\pm$ 0.06 \\
\noalign{\smallskip}
\hline
\hline
\noalign{\smallskip}
\multicolumn{12}{c}{\textbf{\lp8}}\\
\noalign{\smallskip}
\hline
\hline
\noalign{\smallskip}
1 & 1353.4 & 138.52\deg & 4.82$\pm$0.36 & 1 & 1356.2 & 138.64\deg & 4.74 $\pm$0.27& 1 & 1355.0 & 138.71\deg & 4.26$\pm$0.26 \\
2 & 1354.5 & 138.58\deg & 4.57$\pm$0.28 & 2 & 1356.6 & 138.62\deg & 4.63 $\pm$0.34& 2 & 1357.6 & 138.85\deg & 3.86$\pm$0.25 \\
3 & 1354.8 & 138.58\deg & 4.77$\pm$0.30 & 3 & 1355.9 & 138.59\deg & 4.64 $\pm$0.35& 3 & 1357.0 & 138.85\deg & 4.12$\pm$0.26 \\
4 & 1354.6 & 138.45\deg & 4.57$\pm$0.36 & 4 & 1354.9 & 138.67\deg & 4.66 $\pm$0.33& 4 & 1356.4 & 138.68\deg & 4.16$\pm$0.26 \\
5 & 1356.2 & 138.48\deg & 4.66$\pm$0.30 & 5 & 1357.4 & 138.52\deg & 4.81 $\pm$0.35& 5 & 1356.7 & 138.76\deg & 4.11$\pm$0.25 \\
6 & 1354.8 & 138.54\deg & 4.62$\pm$0.33 & 6 & 1356.1 & 138.55\deg & 4.66 $\pm$0.34& 6 & 1356.7 & 138.63\deg & 4.18$\pm$0.26 \\
7 & 1351.0 & 138.57\deg & 4.62$\pm$0.31 & 7 & 1353.0 & 138.55\deg & 4.66 $\pm$0.34& 7 & 1358.2 & 138.84\deg & 4.13$\pm$0.26 \\
8 & 1353.3 & 138.49\deg & 4.64$\pm$0.35 & 8 & 1355.4 & 138.61\deg & 4.52 $\pm$0.42& 8 & 1357.9 & 138.75\deg & 4.08$\pm$0.26 \\
9 & 1354.3 & 138.53\deg & 4.86$\pm$0.36 & 9 & 1353.3 & 138.63\deg & 4.74 $\pm$0.35& 9 & 1359.2 & 138.86\deg & 3.99$\pm$0.27 \\
10 & 1351.9 & 138.62\deg & 4.66$\pm$0.37 & 10 & 1356.2 & 138.56\deg & 4.49 $\pm$0.37& 10 & 1354.0 & 138.59\deg & 4.26$\pm$0.26 \\
11 & 1353.6 & 138.52\deg & 4.76$\pm$0.36 & 11 & 1356.1 & 138.62\deg & 4.51 $\pm$0.35& 11 & 1357.1 & 138.82\deg & 4.13$\pm$0.25 \\
12 & 1354.4 & 138.57\deg & 4.61$\pm$0.22 & 12 & 1353.9 & 138.56\deg & 4.87 $\pm$0.27& 12 & 1357.4 & 138.81\deg & 4.07$\pm$0.25 \\
13 & 1353.1 & 138.51\deg & 4.88$\pm$0.37 & 13 & 1357.0 & 138.59\deg & 4.70 $\pm$0.37& 13 & 1356.2 & 138.75\deg & 4.02$\pm$0.24 \\
14 & 1353.4 & 138.69\deg & 4.76$\pm$0.40 & 14 & 1356.8 & 138.59\deg & 4.68 $\pm$0.38& 14 & 1355.0 & 138.76\deg & 4.06$\pm$0.26 \\
15 & 1349.1 & 138.39\deg & 4.62$\pm$0.37 & 15 & 1353.4 & 138.70\deg & 4.89 $\pm$0.28& 15 & 1356.2 & 138.59\deg & 4.12$\pm$0.26 \\
16 & 1352.8 & 138.44\deg & 4.92$\pm$0.37 & 16 & 1356.2 & 138.58\deg & 4.48 $\pm$0.33& 16 & 1354.6 & 138.70\deg & 4.11$\pm$0.26 \\
17 & 1355.0 & 138.50\deg & 4.87$\pm$0.31 & 17 & 1353.1 & 138.58\deg & 4.68 $\pm$0.33& 17 & 1354.9 & 138.73\deg & 4.16$\pm$0.26 \\
18 & 1356.3 & 138.53\deg & 4.96$\pm$0.38 & 18 & 1351.6 & 138.52\deg & 4.58 $\pm$0.33& 18 & 1354.1 & 138.59\deg & 4.15$\pm$0.27 \\
19 & 1355.0 & 138.49\deg & 4.56$\pm$0.23 & 19 & 1354.9 & 138.62\deg & 4.36 $\pm$0.31& 19 & 1355.2 & 138.63\deg & 4.13$\pm$0.27 \\
20 & 1354.9 & 138.53\deg & 4.86$\pm$0.30 & 20 & 1352.5 & 138.56\deg & 4.58 $\pm$0.34& 20 & 1350.7 & 138.51\deg & 4.07$\pm$0.28 \\
21 & 1352.4 & 138.47\deg & 4.72$\pm$0.31 & 21 & 1355.1 & 138.67\deg & 4.58 $\pm$0.41& 21 & 1354.4 & 138.53\deg & 4.13$\pm$0.28 \\
22 & 1354.5 & 138.57\deg & 5.02$\pm$0.29 & 22 & 1357.8 & 138.58\deg & 4.44 $\pm$0.33& 22 & 1358.4 & 138.68\deg & 4.12$\pm$0.27 \\
23 & 1351.4 & 138.49\deg & 4.82$\pm$0.35 & 23 & 1354.6 & 138.48\deg & 4.30 $\pm$0.33& 23 & 1351.7 & 138.60\deg & 4.15$\pm$0.28 \\
24 & 1354.7 & 138.57\deg & 5.10$\pm$0.29 & 24 & 1356.2 & 138.58\deg & 4.38 $\pm$0.31& 24 & 1354.6 & 138.56\deg & 4.16$\pm$0.25 \\
25 & 1353.9 & 138.44\deg & 4.95$\pm$0.23 & 25 & 1355.2 & 138.60\deg & 4.38 $\pm$0.33& 25 & 1352.5 & 138.64\deg & 4.10$\pm$0.26 \\
26 & 1349.6 & 138.41\deg & 4.56$\pm$0.24 & 26 & 1353.2 & 138.66\deg & 4.35 $\pm$0.40& 26 & 1355.7 & 138.63\deg & 4.12$\pm$0.27 \\
27 & 1352.4 & 138.45\deg & 4.74$\pm$0.29 & 27 & 1356.7 & 138.55\deg & 4.31 $\pm$0.31& 27 & 1355.4 & 138.57\deg & 4.01$\pm$0.25 \\
28 & 1355.2 & 138.48\deg & 4.95$\pm$0.38 & 28 & 1358.1 & 138.59\deg & 4.39 $\pm$0.38& 28 & 1353.0 & 138.63\deg & 4.14$\pm$0.27 \\
29 & 1352.5 & 138.57\deg & 4.64$\pm$0.30 & 29 & 1353.3 & 138.59\deg & 4.47 $\pm$0.35& 29 & 1355.2 & 138.65\deg & 4.14$\pm$0.26 \\
30 & 1352.6 & 138.47\deg & 4.70$\pm$0.43 & 30 & 1352.6 & 138.61\deg & 4.41 $\pm$0.34& 30 & 1354.4 & 138.63\deg & 4.12$\pm$0.21 \\
\noalign{\smallskip}
\hline \hline

\end{tabular}
\end{table*}

\begin{table*}
\center
\caption{\label{tab:photometry} Measured parameters of the two VLM systems \lp1\ and \lp8.}
\setlength{\tabcolsep}{5.5mm}
\begin{tabular}{c|ccc|ccc}
\hline \hline
\noalign{\smallskip}
\multicolumn{1}{c|}{Binaries} & \multicolumn{3}{c|}{\lp1} & \multicolumn{3}{c}{\lp8} \\
\noalign{\smallskip}
\hline 
\noalign{\smallskip}
& 2MASS & \multicolumn{2}{c|}{Individual components} & 2MASS & \multicolumn{2}{c}{Individual components} \\
Parameter & A+B & A & B & A+B & A & B\\
\noalign{\smallskip}
\hline\hline
\noalign{\smallskip}
$J$ & $9.10 \pm 0.02$ & $9.82 \pm 0.04$ & $9.90 \pm 0.04$ & $8.86 \pm 0.02$ & $9.08 \pm 0.03$ & $10.77 \pm 0.04$\\
$H$ & $8.47 \pm 0.02$ & $9.24 \pm 0.03$ & $9.40 \pm 0.03$ & $8.24 \pm 0.04$ & $8.46 \pm 0.04$ & $10.11 \pm 0.05$\\
$K_s$ & $8.21 \pm 0.02$ & $9.01 \pm 0.05$ & $9.08 \pm 0.05$ & $8.00 \pm 0.02$ & $8.24 \pm 0.02$ & $9.77 \pm 0.03$\\
$G^{\dagger}$ & & --- & --- & & $11.818$ & $13.381$\\
\noalign{\smallskip}
$M_{J}$ & & $8.7 \pm 0.4$ & $8.8 \pm 0.4$ & & $6.44 \pm 0.03$ & $8.13 \pm 0.04$\\
$M_{H}$ & & $8.1 \pm 0.4$ & $8.2 \pm 0.4$ & & $5.82 \pm 0.04$ & $7.47 \pm 0.05$\\
$M_{Ks}$ & & $7.8 \pm 0.4$ & $7.9 \pm 0.4$ & & $5.60 \pm 0.02$ & $7.13 \pm 0.03$\\
$M_{G}$ & & --- & --- & & 10.721 & 12.284 \\
\noalign{\smallskip}
$J-K_s$ & $0.81 \pm 0.06$ & $0.82 \pm 0.06$ & $0.89 \pm 0.03$ & $0.87 \pm 0.03$ & $ 0.84\pm 0.03$ & $1.00 \pm 0.05$ \\
\noalign{\smallskip}
$\Delta J$ && \multicolumn{2}{c|}{$0.08 \pm 0.06$}&& \multicolumn{2}{c}{$1.69 \pm 0.03$}\\
$\Delta H$ && \multicolumn{2}{c|}{$0.16 \pm 0.04$}&& \multicolumn{2}{c}{$1.65 \pm 0.04$}\\
$\Delta K_s$ && \multicolumn{2}{c|}{$0.07 \pm 0.06$}&& \multicolumn{2}{c}{$1.54 \pm 0.02$}\\
\noalign{\smallskip}
SpT$^{\dagger \dagger}$ & M4 & M4.5$\pm$0.5 & M4.5$\pm$0.5 & M4 & M1.0$\pm$0.3 & M3.8$\pm$0.3\\
\noalign{\smallskip}
Position Angle && \multicolumn{2}{c|}{334.0\deg~$\pm$ 0.1\deg} && \multicolumn{2}{c}{138.6\deg~$\pm$ 0.1\deg}\\
Separation [mas] && \multicolumn{2}{c|}{$402 \pm 1$} && \multicolumn{2}{c}{1355 $\pm$ 2}\\
\noalign{\smallskip}
\hline\hline
\end{tabular}
\\[0mm]
$^{\dagger}$ -- \textit{Gaia} DR2 magnitude; $^{\dagger \dagger}$ -- Spectral type of combined system is adopted from \cite{Rajpurohit-13-A+A-2}. For the individual systems, the spectral type have been derived from $M_{J}$ using the method of \cite{Scholz-05-A+A-9}.
\end{table*} 


\section{Data Reduction and analysis}
\label{sec:datared}
To carry out our analysis, we have used the raw image mode data that were provided in ESO archive\footnote{\href{https://archive.eso.org/eso/eso_archive_main.html}{https://archive.eso.org/eso/eso\_archive\_main.html}}.
The preliminary data reduction were performed using the \textit{Image Reduction and Analysis Facility} \citep[\iraf\footnote{\href{http://iraf.net}{http://iraf.net}}; ][]{TodyD-86-SPIE, TodyD-93-ASPC} software. We have carried out the dark subtraction, flat fielding, sky subtraction using each of the jittered images, and cosmic ray removal using the standard packages available in \iraf.

A close-up view of the representative $J$-band images for each of the objects \lp1\ and \lp8\ are shown in Fig.~\ref{fig:Image} where the usual conventions were followed, i.e. North is up, and East is left. Both \lp1\ and \lp8\ are clearly resolved as a binary with two components. In the case of \lp1, the components have almost equal brightness with a slightly higher flux of the South-East component (\lp1~A) than the North-West component (\lp1~B). However, \lp8\ consists of two components having very different brightness, where the North-West component (\lp8~A) is clearly observable as the brighter one than the South-East component (\lp8~B). In our analysis, we have taken the brighter component as primary. 

We have measured the CCD chip coordinates (i.e. positions) of the sources on each of the science images using the Python task {\sc daostarfinder} \citep[][]{Stetson-87-PASP-3} available in {\sc photutils} package. The task {\sc daostarfinder} finds the object centroid by fitting the marginal x and y one-dimensional distributions of Gaussian kernel to the marginal x and y distributions of the unconvolved image.
The separation between the centroids of the primary and secondary components was derived for each of the science images. The derived separations for both the objects in all $JHK_s$ images are given in the 2$^{nd}$, 6$^{th}$, and 10$^{th}$ column of Table~\ref{tab:measurements}.
The binary separations are found to be consistent for both the sources. For \lp1\ we derive a separation of 402$\pm$1, 401$\pm$1, and 401$\pm$1 mas for $JHK_s$ photometric bands, whereas for \lp8\ these values are calculated to be 1354$\pm$2, 1355$\pm$2, and 1356$\pm$2 mas, respectively. 
Using the python package {\sc astropy}, we have converted all the pixel coordinates to World Coordinate System (WCS), whereas in order to derive the position angles (PA) of each of the images we have used the {\sc pyastronomy} package. 
The derived values of PA of the secondary with respect to the primary components are listed in 3$^{rd}$, 7$^{th}$, and 11$^{th}$ column of Table~\ref{tab:measurements} for $J$, $H$, and $K_s$ bands, respectively. These values show consistency for both the sources. The calculated values of PA for \lp1\ are 333.99\deg$\pm$0.16\deg, 333.95\deg$\pm$0.10\deg, and 334.07\deg$\pm$0.10\deg\ for $JHK_s$ photometric bands, whereas in case of \lp8\ the PA were derived to be 138.52\deg$\pm$0.06\deg, 138.60\deg$\pm$0.05\deg, and 138.68$\pm$0.10\deg, respectively. 

In order to derive the flux ratio and the instrumental magnitude, aperture photometry was performed on both the components in each frame for both the sources \lp1\ and \lp8. 
Using the \iraf\ task {\sc phot} we properly centre the source, and with the given radius of the aperture as input, it derives the background-subtracted sum of all the photons within that aperture. This task also allows us to specify a series of the increasing aperture from an optimal small aperture (FWHM of the source profile) to a larger aperture (10 times of the FWHM). The increment of the aperture was 0.5 pixel. The larger the aperture, the more the flux of the source would be enclosed by the aperture. However, with a larger aperture, the errors introduced due to sky subtraction would also be larger. 

Theoretically, a stellar (Gaussian) profile extends up to infinity. Therefore, the magnitude which is restricted for the chosen aperture needs to be corrected for the excess counts in the wings of the stellar profile. The correction from profile fitting magnitude to aperture magnitude is carried out by the process of determining the aperture growth curve, i.e. a plot of magnitude within a given aperture versus aperture size. The aperture correction is simply the magnitude difference between the asymptotic magnitude and the magnitude at the given aperture. The most straightforward way to determine the aperture correction is to measure it directly from a number of growth curves. 
A more advanced method for doing the aperture correction is the {\sc daogrow} algorithm \citep{Stetson-90-PASP-3}, implemented in the \iraf\ task {\sc mkapfile} in the package {\sc digiphot.photcal}. Using this task, a stellar profile model was fitted to the growth curves for one or more components in one or more images. It then computes the aperture correction from a given aperture to the largest aperture. 
Again, we also took great care that the neighbouring components would not start influencing the sky background of the selected component. In order to do that we have plotted the intensity variation along the axis (line joining the centroid of both components). We have taken the minimum intensity of this plot as the upper-limit of aperture. In each frame, we again verified manually to avoid any unwanted mistake, and the optimal aperture is selected.


\section{Results}
\subsection{NIR Photometry }
\label{sec:nir_photometry}
Using the procedure described in section~\ref{sec:datared}, we have estimated the instrumental magnitudes which were then standardised to apparent magnitude using the zero-point images that were routinely taken for NaCo service operation. The results for individual $J$, $H$, and $K_s$ bands apparent magnitudes are summarised in the first three rows of Table\,\ref{tab:photometry}. We have also estimated the flux ratio as the ratio of the flux of primary to secondary components in each of the frames. The results are shown in 4$^{th}$, 8$^{th}$, and 12$^{th}$ column of Table~\ref{tab:measurements} for $J$, $H$, and $K_s$ bands, respectively. The uncertainties in flux ratios that are shown in the table were estimated using the propagation of the photometric uncertainty. 

The photometric measurements for the unresolved combined system of \lp1\ and \lp8\ in $J$, $H$, and $K_s$ bands were available from the 2MASS data. The 2MASS magnitudes of \lp1\ and \lp8\ reflect the combined flux from both components. In order to carry out a more robust analysis, we have also derived the magnitude of the individual components from the combined flux and the flux ratio as derived in Table~\ref{tab:measurements}. 
We found that the apparent magnitudes derived in this method are consistent with our photometric analysis within 2$\sigma$.
In the case of \lp8, the detection of the components of \lp8~A and B, confirms the observations from \textit{Gaia} DR2 \citep[][]{Bailer-Jones-18-AJ-6}. The binary separation and position angle derived  in this work also matches with the results from \textit{Gaia}, however, previous authors did not calculate detailed astrometry. The individual \textit{Gaia} magnitudes of \lp8~AB were $G$ = 11.82 and 13.38, respectively, for the A and B component.

As discussed in Section~\ref{sec:intro}, following \cite{Finch-14-AJ-8} and \cite{Bailer-Jones-18-AJ-6}, in this work we have adopted the distances for \lp1\ and \lp8\ as 16.57$\pm$3.27 and 33.77$\pm$0.17 pc, respectively. 
In order to assign the above-measured distances as the distance of individual spatially separated components (A and B) of \lp1~AB and \lp8~AB, it is important to be sure that the A and B components really comprise a true binary system, rather than an unrelated pair with a small projected separation. 
In our analysis, the $J - K_s$ colour and derived spectral-type for both the components in each of the systems \lp1~AB and \lp8~AB are found to have similar values within 1--2$\sigma$ (see 9$^{th}$ row in Table~\ref{tab:photometry}). Since both the systems share the equal flux ratio and colour between the two individual components, the chance of either of the A or B component of the system for being a background object is very little and can be ruled out \citep[][]{Dahn-02-AJ-2, Hawley-02-AJ-5, West-08-AJ-7, Bonfini-09-A+A}.

With the assumption that the two components (A and B) in each of the observed systems \lp1~AB and \lp8~AB are indeed part of the physical binary systems (i.e., both components in each system are located at the same distance) the absolute magnitudes of the individual components can be determined for each individual components of the binary systems.
The derived absolute magnitudes ($M_{J}$, $M_{H}$, $M_{Ks}$, and $M_{G}$) for both the systems in $J$, $H$, $K_s$, and $G$ (only for \lp8) photometric bands are given in 5$^{th}$--8$^{th}$ row of Table~\ref{tab:photometry}. Due to the larger uncertainty in distance measurement for \lp1~AB, the estimated uncertainties in the absolute magnitude of \lp1~AB are derived to be approximately ten times larger than the uncertainties in \lp8~AB. 

\subsection{Spectral Classification}
\label{sec:sptype}

\begin{figure}
\centering
\subfloat{\includegraphics[height=6.6cm, trim=0 0 1cm 0,clip]{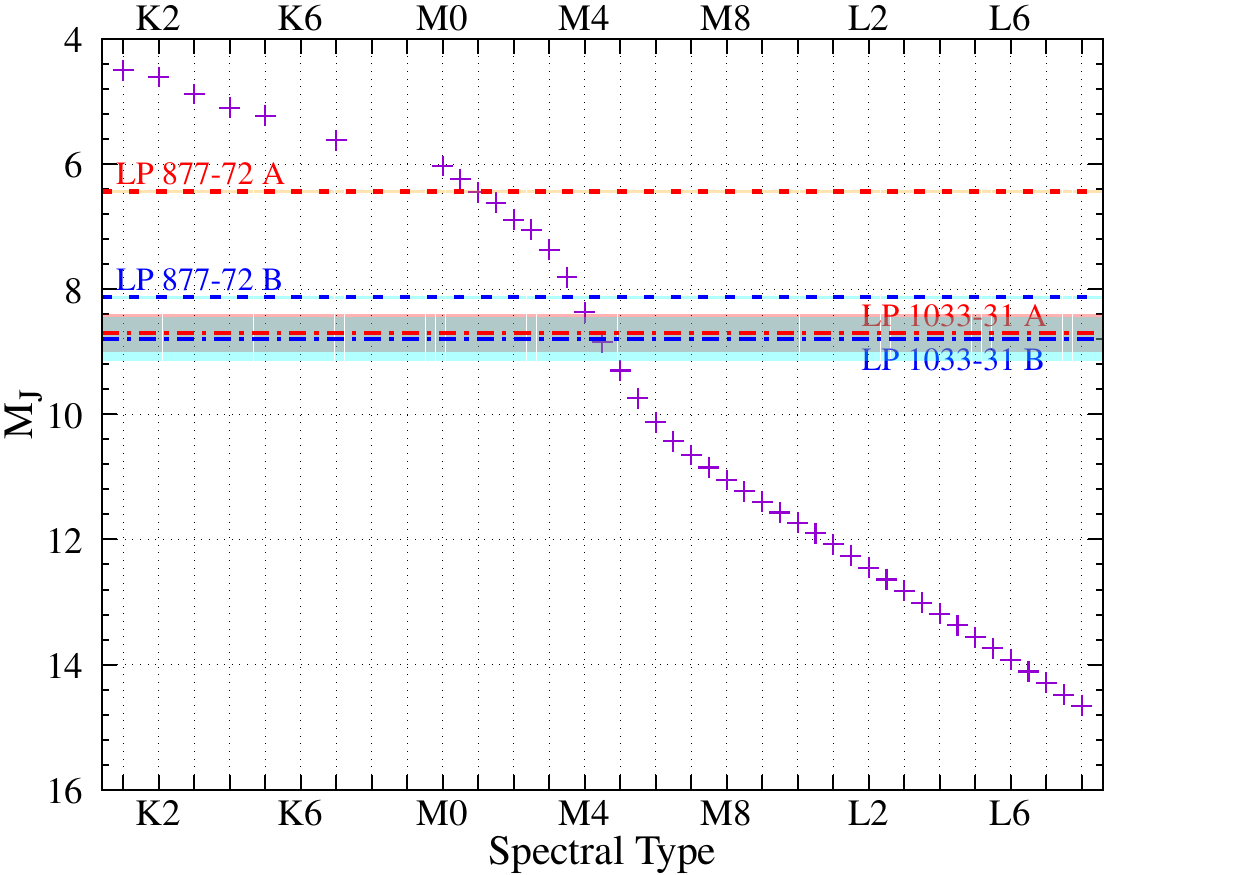}}
\caption{Absolute magnitude $M_J$ as a function of spectral types adopted from \pcite{Scholz-05-A+A-9} is shown by `+' symbol. The dashed and dot-dashed lines show the derived $M_J$ of \lp8~AB and \lp1~AB, respectively. The red and blue colours denote, respectively, the primary (A) and secondary (B) component of each system. The orange and cyan shades shows the respective uncertainties in A and B components.}
\label{fig:sp-type-scholz}
\end{figure}

In recent years, several empirical relationships have been derived between the absolute $J$-magnitude ($M_{J}$) and spectral type of the stars \citep[e.g.][]{Dahn-02-AJ-2, Cruz-03-AJ-4, Scholz-05-A+A-9}. These relationships are very useful in order to estimate the spectral types of the individual stellar components of the VLM stars. In our analysis, we have used the new calibration from the empirical data of \cite{Scholz-05-A+A-9} to determine the spectral type. This is because unlike other methods, this relationship is applicable within a wide spectral range of K0--L8. Fig.~\ref{fig:sp-type-scholz} shows the relation of the spectral type and absolute magnitude $M_{J}$ adopted from \cite{Scholz-05-A+A-9} along with the derived $M_{J}$ of the individual components of \lp1~AB and \lp8~AB with dashed and dot-dashed lines. We found that the spectral type of both \lp1~A and B are similar within their uncertainty range and derived to be M4.5$\pm$0.5. In the case of \lp8~A and B components, the derived spectral types are M1.0$\pm$0.3 and M3.8$\pm$0.3, respectively. The uncertainty in each case of spectral type was derived propagating the uncertainty of all involved parameters.

\section{Physical Parameters of VLM Binaries}
\label{sec:parameters}
\subsection{Ages}
\label{subsec:ages}
Estimating the exact age for any of these binaries \lp1\ and \lp8\ are difficult since there are no Li measurements yet published \citep[which could place an upper limit on the ages; see][]{Zboril.-97-MNRAS-4}.
Using the Hipparcos data, \cite{CaloiV-99-A+A-1} have estimated the age of stellar populations in the solar neighbourhood (0.6--7.5 Gyr) and categorised the sample with respect to their velocity. Using the proper motion and distances as mentioned in Section~\ref{sec:intro}, for \lp1\ and \lp8, we have estimated the tangential velocity of 23.48 and 48.82 km s$^{-1}$, respectively. This sets a lower limit to the age range of \lp1\ and \lp8\ binary systems to be $\geq$0.01 and $\geq$1.5--2 Gyr, respectively \citep[][]{CaloiV-99-A+A-1}.

Since both the binary systems consist of M-dwarfs, we can get a crude estimation of the age of the individual components using the age-activity relation for M-dwarfs \citep{West-08-AJ-7}. Initially, considering the combined systems, both objects \lp1\ and \lp8\ have combined spectral type M4 \citep[using 2MASS data;][]{Rajpurohit-13-A+A-2}, which corresponds to an age range of 4.5$_{-1.0}^{+0.5}$ Gyr from the age-activity relation \citep{West-08-AJ-7}. However, considering the individual systems, the estimated age for both the primary and secondary components of \lp1\ would lie within the range of 3.5--7.5 Gyr or $\approx$5 Gyr. In case of \lp8, the individual components give quite different ages for primary and secondary ($\lsimeq$ 0.8 and 1.5--5.0 Gyr) which is unlikely for a physically bound system since both components of the binary system supposed to be created at the same age. Using the evolutionary calculations by \citep[][]{ChabrierG-97-A+A-2, ChabrierG-00-ApJ-3}, the typical age for the mass range of $\gsimeq$ 0.15 \msun\ were estimated to be $<$ 10 Gyr \citep[][]{ChabrierG-97-A+A-2, ChabrierG-00-ApJ-3}. It should be noted that there is little significant difference between the evolutionary tracks for ages 0.6--10 Gyr for these spectral types. Therefore, we have conservatively assumed that age of $\sim$5 Gyr.

\begin{table}
\center
\small
\caption{\label{tab:parameters} System parameters for \lp1\ and \lp8\ assuming the age of 5 Gyr}
\setlength{\tabcolsep}{0.85mm}
\begin{tabular}{l|cc|cc}
\hline
\hline
\noalign{\smallskip}
\multirow{2}{*}{Par$^{\dagger\dagger}$} & \multicolumn{2}{c|}{\lp1}& \multicolumn{2}{c}{\lp8} \\
& A & B & A & B\\

\noalign{\smallskip}
\hline
\hline
\noalign{\smallskip}
$d$ & \multicolumn{2}{c|}{16.57$\pm$3.27} & \multicolumn{2}{c}{33.77$\pm$0.17} \\[0.4mm]
a & \multicolumn{2}{c|}{6.7$\pm$1.3} & \multicolumn{2}{c}{45.8$\pm$0.3} \\[0.4mm]
M & 0.20$\pm$0.04 & 0.19$\pm$0.04 & 0.520$\pm$0.006 & 0.260$\pm$0.005 \\[0.4mm]
R & 0.225$\pm$0.030 & 0.217$\pm$0.029 & 0.492$\pm$0.011 & 0.270$\pm$0.006 \\[0.4mm]
$\teff$ & 3245$\pm$107 & 3211$\pm$120 & 3750$\pm$15 & 3348$\pm$10 \\[0.4mm]
$\logg$ & 5.061$\pm$0.038 & 5.068$\pm$0.039 & 4.768$\pm$0.005 & 4.984$\pm$0.004 \\[0.4mm]
L & 0.0050$\pm$0.0015& 0.0045$\pm$0.0014 & 0.0432$\pm$0.0021 & 0.0082$\pm$0.0004 \\[0.4mm]
P & \multicolumn{2}{c|}{28.0$\pm$2.6} & \multicolumn{2}{c}{349.4$\pm$2.5} \\[0.4mm]
E$_{Bind}$ & \multicolumn{2}{c|}{-10.0$\pm$3.5} & \multicolumn{2}{c}{-5.21$\pm$0.12} \\[0.4mm]
\noalign{\smallskip}
\hline
\hline
\end{tabular}
\\[1mm]
\scriptsize
$^{\dagger\dagger}$ -- The parameters in first column: 
$d$ -- Distance in pc; a -- Separation in AU; M -- Mass in \msun\ as derived from the interpolation of \cite{BaraffeI-15-A+A} isochrones; R -- Radius in \rsun; $\teff$ -- Effective Temperature in K; $\logg$ -- Logarithmic value of surface gravity in cgs unit; L -- Luminosity in \lsun; P -- Orbital period in yr; E$_{Bind}$ -- Binding Energy in 10$^{43}$ erg ;\\
\normalsize
\end{table}
\subsection{Masses, Surface gravities, and Effective Temperatures}
\label{subsec:mass_logg_teff}
Initially, we have estimated the masses of both the components of \lp1\ and \lp8\ using the mass-absolute magnitude relation as derived by \cite{BenedictG-16-AJ}. From the K-band absolute magnitudes as derived in Section~\ref{sec:nir_photometry}, we have estimated the mass of \lp1~A and B to be 0.19$\pm$0.04 and 0.18$\pm$0.04~\msun, respectively. In case of \lp8, the respective values of masses of the primary and secondary are derived to be 0.565$\pm$0.004 and 0.286$\pm$0.005~\msun. 
In order to carry out a more robust analysis, we have also interpolated the \cite{BaraffeI-15-A+A} isochrones for an age of 5 Gyr. \lp1~AB are found to have the nearly equal mass of 0.20$\pm$0.04 and 0.19$\pm$0.04~\msun, whereas, for \lp8~AB, the derived masses are found to be 0.520$\pm$0.006 and 0.260$\pm$0.005~\msun, respectively. Although both the methods of estimating masses gives similar results within 3$\sigma$ and 5$\sigma$ for \lp1\ and \lp8, we have used the later method of mass estimation for further analysis in this paper due to its robustness. The derived mass, $\teff$, and $\logg$ are given in Table~\ref{tab:parameters}. For \lp1, the $\teff$ and $\logg$ are found to be similar, whereas in the case of \lp8, the primary is $\sim$400 K hotter and $\logg$ is $\sim$1.6 times less than the secondary component.

\subsection{Stellar Radii and Luminosities}
\label{subsec:radius_lumin}
In order to derive the stellar radius for both the stars, we have used single-star mass-radius relations \citep[][]{Boyajian-12-ApJ-3}. The radii of LP 1033-31 A and B are derived to be 0.225$\pm$0.030 and 0.217$\pm$0.029 \rsun. Whereas for LP 877-72 AB, the radius is estimated to be 0.492$\pm$0.011 and 0.270$\pm$0.006 \rsun, for primary and secondary respectively.
Using the Stefan-Boltzmann law, we have estimated the luminosity of both the primary and secondary components of \lp1\ and \lp8\ (see Table~\ref{tab:parameters}). Further, we have also derived the luminosity using the Stellar Luminosity versus Temperature relation given by \cite{Boyajian-12-ApJ-3}. Both the methods are found to be consistent within its uncertainty level. 

\subsection{Binary Separations, Orbital Periods, and Binding Energies}
\label{subsec:separation_period_binding_energy}
In Section~\ref{sec:datared}, we have derived the projected angular separation between the individual components (between A and B) of \lp1\ and \lp8\ to be 402$\pm$1 and 1355$\pm$2 mas. Incorporating the distances taken from \cite{Bailer-Jones-18-AJ-6} and \cite{Finch-14-AJ-8}, as discussed in Section~\ref{sec:intro}, we derived the projected physical separation between the two systems of 6.7$\pm$1.3 and 45.8$\pm$0.3 AU, respectively. Because of the projection effects, the real separation between the components is expected to be, on average, 1.4 times larger \citep[][]{CouteauP-60-JO-1}. However, in this study, we have taken the measured projected separations, and the masses derived in Section~\ref{subsec:mass_logg_teff} to estimate the orbital periods for each of the systems. The orbital period derived for \lp1\ is found to be 28$\pm$3 yrs whereas for \lp8\ it is derived to be 349$\pm$3 yrs. Therefore, none of our candidates will have measurable orbital motions and hence will have a common proper motion on the sky.

\begin{figure*}
\centering
\subfloat[LP~1033-31]{\includegraphics[height=8.8cm, angle=-90]{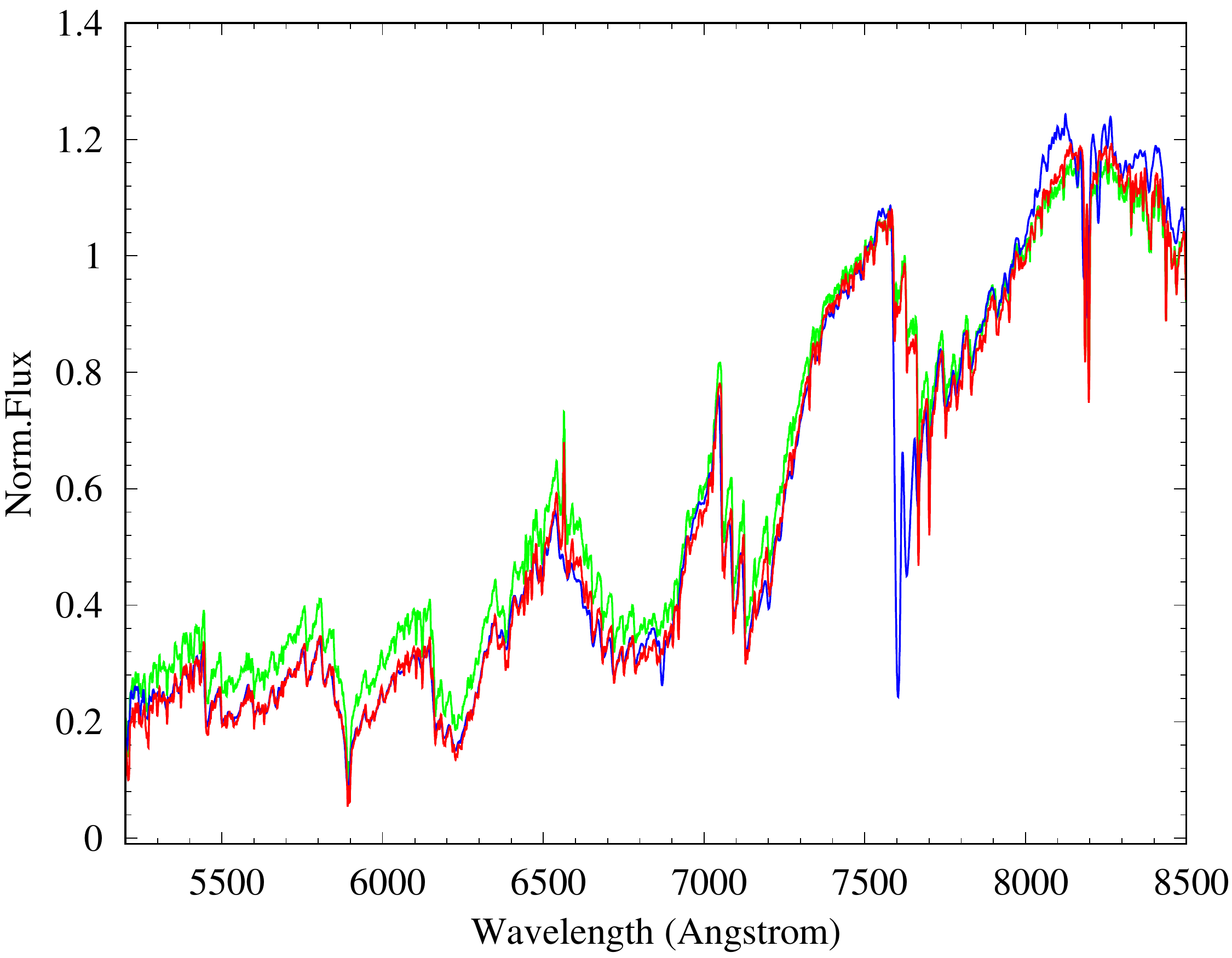}}
\hspace{0.2mm}
\subfloat[LP~877-72]{\includegraphics[height=8.8cm, angle=-90]{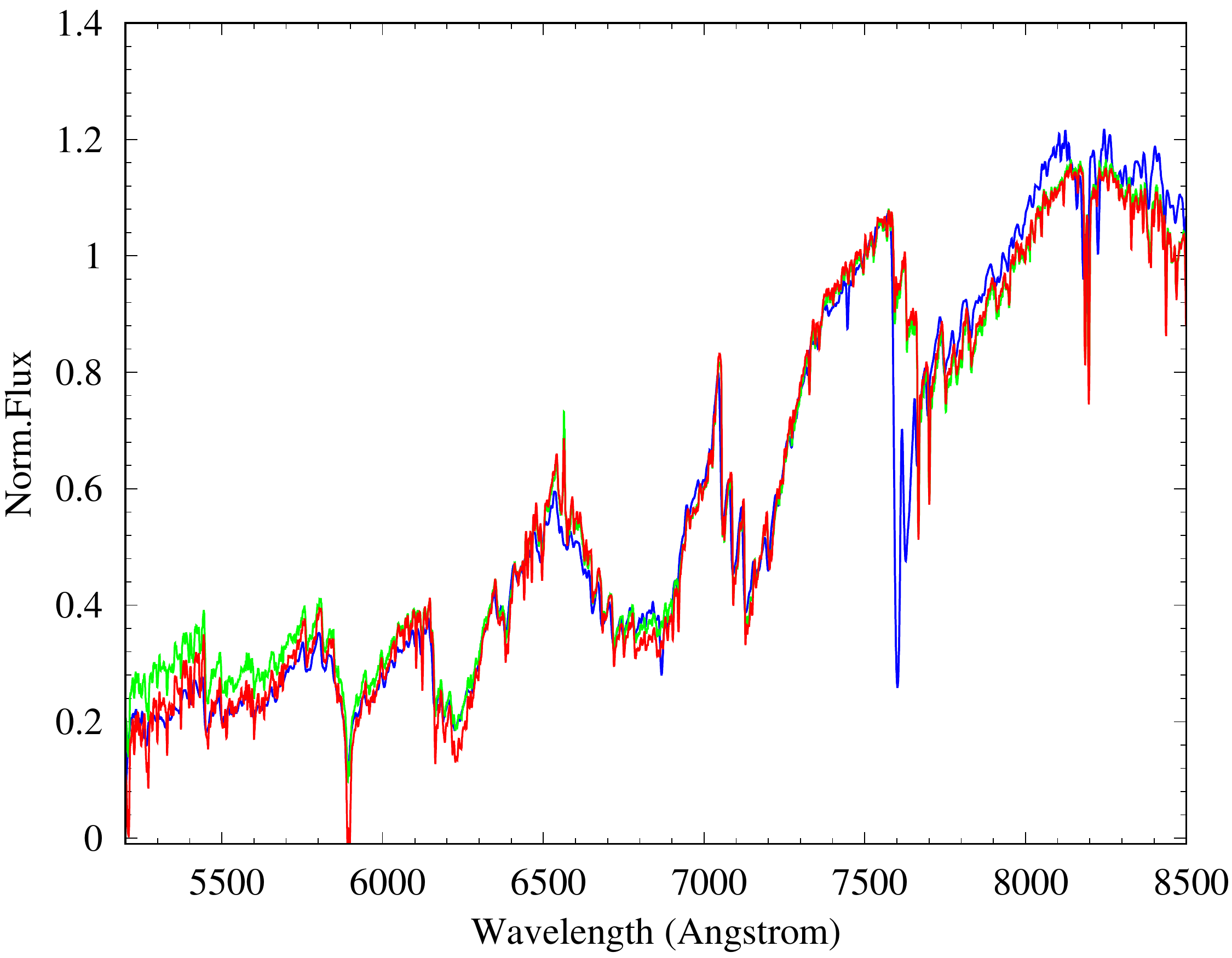}}
\caption{Combined template fit to the observed spectra. The observed 2MASS spectra of the unresolved systems \lp1\ (a) and \lp8\ (b), adopted from \pcite{Reyle-06-MNRAS-3}, are shown with blue solid line. The best-fit spectra created with the combination of the templates of the primary and secondary of each systems using the derived spectral types from our analysis (see Section~\ref{sec:sptype}) and in relevant flux ratios (as given in Table~\ref{tab:measurements}) have been shown with red solid line. We have adopted the latest template spectra as given by \pcite{Kesseli.-17-ApJS-2}. For comparison, we have also shown the best-fit template spectra for the previously unresolved systems (green solid line) using the spectral types derived by \pcite{Rajpurohit-13-A+A-2} (as mentioned in Table~\ref{tab:photometry}). This is clearly evident that the combined template fitting using the spectral type of the individual components from this study shows a better fit to the observed spectra than the spectra of the unresolved system.
}
\label{fig:spectra}
\end{figure*}

Since low mass binary systems are expected to have low (absolute values of) gravitational potential (binding) energies, in order to understand if these systems are physically bound or not, we have calculated the binding energies for \lp1\ and \lp8. Using the masses and projected physical separations (instead of the true separations), the binding energies ($U$~=~--G.M$_{1}$.M$_{2}$/r; where M$_{1}$ and M$_{2}$ are the masses of the components and r is the distance between them) for \lp1\ and \lp8\ are derived to be --1.00$\pm$0.35~$\times$~10$^{44}$ and --5.21$\pm$0.12~$\times$~10$^{43}$ erg, respectively. We have further discussed the stability of the systems in Section~\ref{subsec:disc_stability}.

\section{Discussion}
\label{sec:discussion}
Using high-resolution NaCo/VLT images, we have detected and characterised two VLM binaries \lp1\ and \lp8. In this section, we have discussed how the understanding of these two sources has been changed through our study. We also discussed whether the identified components, in this study, match with other similar kinds of systems or not. 

\subsection{The detected binaries: previous and present understanding}
\label{subsec:disc_properties}
In Section~\ref{sec:sptype}, we characterised both \lp1\ and \lp8\ to be the VLM binaries, and estimated the masses of each component of \lp1\ to be $\sim$0.2 \msun, whereas for primary and secondary of \lp8, the derived values of masses are 0.52 and 0.26 \msun. Using TIC catalouge, the latest values of masses for \lp1\ and \lp8\ as `single' objects were estimated to be 0.316 and 0.600 \msun\ \citep[][]{StassunK-19-AJ-2, CloutierR-19-AJ-5}. Both of these values were found to be greater than masses of the individual components, however slightly lower (within few $\sigma$ uncertainty level) than the total estimated masses (M$_A$+M$_{B}$) in our study. 

The radii of both \lp1\ and \lp8\ were estimated by \cite{StassunK-19-AJ-2} to be 0.319 and 0.556 \rsun, respectively. These values were larger than both of the primary and secondary components of each VLM binaries. Considering `single' object, the latest values of $\logg$ and $\teff$ were estimated to be 4.93 and 3368 K for \lp1\ \citep[][]{StassunK-18-AJ-2, Muirhead.-18-AJ-3}, and 4.63 and 3782 K for \lp8\ \citep[][]{Anders.-19-A+A-2, Gaia-Collaboration-18-A+A-5}. In our analysis, we found that $\logg$ and $\teff$ for both the newly discovered components of \lp1\ are within $\sim$1.5--3 $\sigma$ of that derived in earlier studies. For \lp8, we found that the previously derived values of surface gravity and temperatures are more close to the brighter component (i.e. the primary). This could be due to the fact that the secondary component is 5.2 times less luminous and 1.53--1.69 mag less bright than the primary, and hence the contribution of the secondary component to the total flux is only $\sim$20\%.

\subsection{Combined Spectral fitting}
\label{subsec:disc_spectral_fit}
Since the spectral type for each of the binary components of both the sources (\lp1\ and \lp8) are found to be different than those were estimated in previous studies, we have performed a more robust analysis. Although we do not have the spectra for each of the individual components, the spectra of the  unresolved systems for both \lp1\ and \lp8\ were already been observed and studied by \cite{Reyle-06-MNRAS-3}. We have used the latest available empirical template library of stellar spectra from \cite{Kesseli.-17-ApJS-2}, and fitted the combined spectra at the flux ratios derived in the Section~\ref{sec:datared}. Since the empirical template library of stellar spectra does not provide any fractional spectral class, we have combined the spectral types to estimate it from the nearest spectral classes (e.g. the spectra of spectral class M4.5 is derived averaging the spectral class M4 and M5).

In Fig.~\ref{fig:spectra}, the combined spectra of \lp1\ (left) and \lp8\ (right) are shown by blue solid lines that were observed by \cite{Reyle-06-MNRAS-3}. Whereas the red solid line in each plot shows the combined template spectra of the primary and secondary using the derived spectral types in this study. In order to compare whether the derived spectral class, from our analysis, give any better fit or not, we have also overplotted the best-fit template spectra of the unresolved systems (green solid line). We have used the spectral class from \cite{Rajpurohit-13-A+A-2} as shown in Table~\ref{tab:photometry}. It is clearly evident that the fitting is better when the spectra of the individual components are combined. This might be due to the better understanding of the system after both the components get spatially resolved.

\subsection{Mass Ratios}
\label{subsec:mass_ratio}

Investigation of the binary mass ratios is very important since the mass-ratio distributions provide stringent tests for models of binary star formation \citep[see][and references therein]{El-BadryK-19-MNRAS-2}. Most studies of binary stars fit a single power-law distribution f$_{q}~\!\!\propto~\!\!q^{\gamma}$ to the observed mass ratios \citep[][]{ShatskyN-02-A+A, SanaH-12-Sci-4, DucheneG-13-ARA+A-1, De-RosaR-14-MNRAS-5}. However, with large samples, it was noticed that a single-parameter model could not adequately fit the distribution across all mass ratios \citep[][]{DuquennoyA-91-A+A-1, HalbwachsJ-03-A+A-2, GulliksonK-16-AJ-3, MoeM-17-ApJS-2, MurphyS-18-MNRAS-4, El-BadryK-19-MNRAS-2}, and hence a multi-parameter power-law model was adopted with different slopes (such as $\gamma_{small-q}$ and $\gamma_{large-q}$) across the small to large mass ratios. The values of the slopes also depend on the projected separations of the binaries as well as the spectral types of the individual components. Therefore, in order to provide observational constraints on the theoretical models, observations of mass ratios are necessary. 

\begin{figure}
\centering
\includegraphics[height=8.4cm, angle=-90]{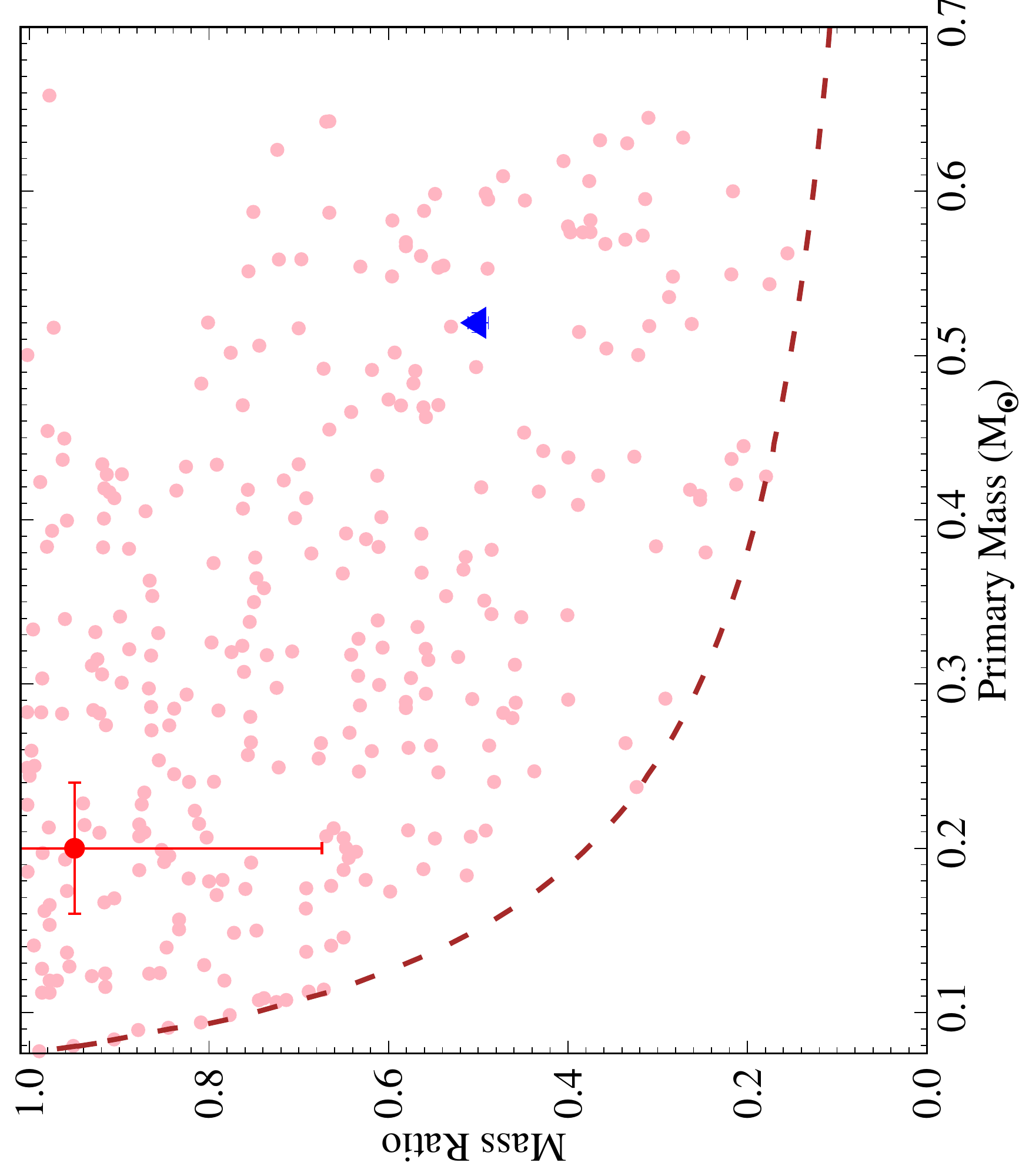}
\caption{
Mass ratio as a function of Primary Mass for \lp1~AB and \lp8~AB have been shown with red solid circle and blue solid triangle, respectively. For comparison with other M-dwarf binaries, we have overplotted the data from the latest work of \pcite{Winters-19-AJ-4} with light-pink solid circle. The brown dashed line indicates the mass ratio boundary relative to the lowest mass considered in their survey (i.e the Hydrogen burning limit M $\gsimeq$ 0.075 \msun). 
}
\label{fig:mass_ratio}
\end{figure}

In this work, we have derived the mass ratios of \lp1\ and \lp8\ to be 0.95$\pm$0.28 and 0.50$\pm$0.01, respectively. In a study of multiplicity among M-dwarfs, \cite{FischerD-92-ApJ-4} found that the distribution of mass ratios is flat and perhaps declines at low-mass ratios. By analysing a catalogue of 1342 low-mass binaries with at least one mid-K to mid-M dwarf component, \cite{DhitalS-10-AJ-1} found the distribution of mass ratios to be strongly skewed toward equal-mass pairs: 85.5\% of pairs have masses within 50\% of each other. A similar result is obtained by \cite{ReidI-97-AJ-1} in the volume-complete 8~pc sample, which includes mostly (80\%) of M-dwarfs. In recent years, M-dwarf binaries are weighted toward more equal masses \citep[see][for a review]{Winters-19-AJ-4}. Figure~\ref{fig:mass_ratio} shows the mass ratios of both the objects (\lp1\ and \lp8) with respect to the mass of the primary component. For a comparison with M-dwarfs, we have overplotted all the pairs in the sample taken by \cite{Winters-19-AJ-4}. 
For early M-dwarf binaries with intermediate separations a $\approx$ 10 AU, the mass-ratio distribution is nearly uniform with a turnover in the sub-stellar brown dwarf regime \citep[][]{FischerD-92-ApJ-4, BergforsC-10-A+A-2, JansonM-12-ApJ-3, Winters-19-AJ-4}. 
For binaries with late M-dwarf primaries and intermediate separations a $\approx$ 1--10 AU, the mass-ratio distribution is weighted significantly toward q $\gsimeq$ 0.7 \citep[][]{BouyH-03-AJ-1, JoergensV-06-A+A-2, BasriG-06-AJ-2, BergforsC-10-A+A-2, DieterichS-12-AJ-2, DucheneG-13-ARA+A-1, Winters-19-AJ-4}. In this study, we also noticed a similar feature for \lp1, which having both components of spectral type M4.5, shows similar mass range. However, with an early-M-dwarf primary component, \lp8\ shows a mass ratio of $\sim$0.5.

\begin{figure}
\centering
\includegraphics[width=8.4cm, angle=0, trim=0 0 3.95cm 0,clip]{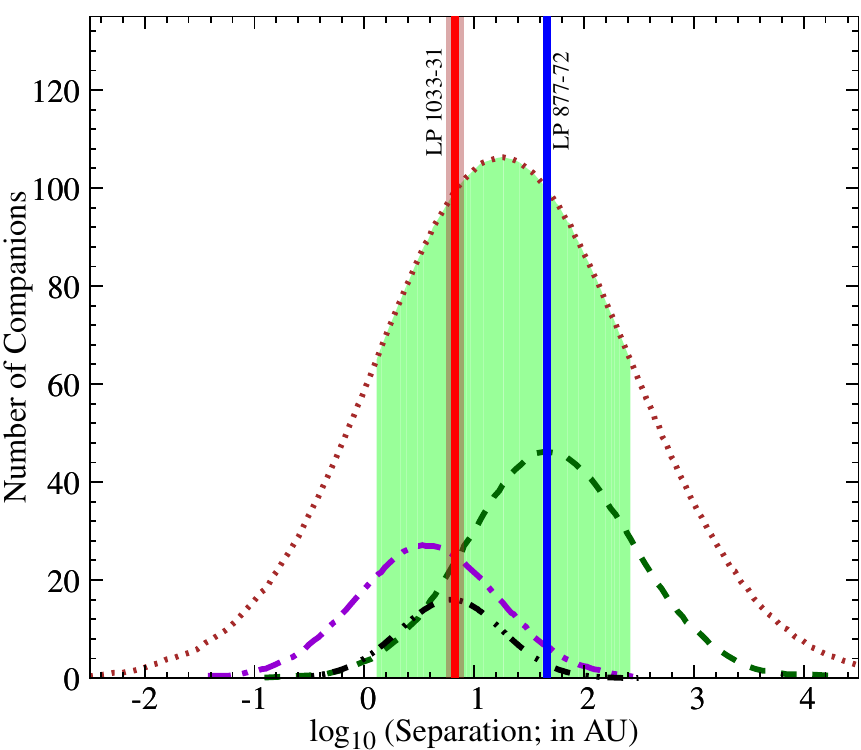}
\caption{
The projected separations of the binaries \lp1~AB and \lp8~AB have been shown with the red and blue solid vertical lines, respectively. The semi-transparent shades with same colors show the respective estimated uncertainties. The uncertainties for \lp1~AB are clearly noticeable, whereas for \lp8~AB, the extent of uncertainties are smaller than the size of the line width.
 For comparison, the separation distributions of early- and mid-M-dwarf binaries estimated by \pcite{JansonM-14-ApJ-4} have been shown by black dot-dot-dashed line and has a peak at $\sim$6 AU. For a larger sample of M-dwarf binaries, the distribution within 10 pc and 25 pc estimated by \pcite{Winters-19-AJ-4} and have been shown with violet dot-dashed line (peak $\sim$4 AU) and brown dotted line (peak $\sim$20 AU), respectively. The semi-transparent green shaded region under the brow dotted line indicates the $\sigma_{log~a}$~=~1.16, for the sample within 25 pc, as adopted from \pcite{Winters-19-AJ-4}. The dark-green dashed line shows the fit for solar-type stars from \pcite{RaghavanD-10-ApJS-2}, which peaks at $\sim$51 AU.
}
\label{fig:sep_dist}
\end{figure}

\subsection{Binary Separation}
\label{subsec:disc_separation}

In this work, we have estimated the binary separation of the binaries \lp1~AB and \lp8~AB to be 6.7$\pm$1.3 and 45.8$\pm$0.03 AU, respectively (see Section~\ref{subsec:separation_period_binding_energy}). In a series of studies in solar neighbourhood M-dwarfs and VLM binaries \cite{FischerD-92-ApJ-4}, \cite{Close-03-ApJ-4}, \cite{GizisJ-03-AJ-5}, and \cite{SieglerN-05-ApJ-2} and few other authors have shown that the separation of the binary systems lies within a maximum separation of $\sim$30 AU. \cite{Close-03-ApJ-4} have even investigated the reason behind the lack of wide (a $>$ 20 AU, which was the observational upper-limit of binary separation in their survey) VLM/brown dwarf binaries. They have suggested a possible significant differential velocity kick, as was predicted by the ``embryo ejection'' theories. Their estimates also indicated a fragmentation-produced VLM binary semi-major axis distribution contains a significant fraction of wide VLM binaries in contrast to their observation. 

The first known VLM binary wider than 30 AU were discovered by \cite{Luhman-04-ApJ-6} with a projected separation a~$\sim$~240 AU. In the later studies, more and more wide VLM binaries have been discovered and analysed by several authors which includes the work of \cite{Chauvin-04-A+A-3} (a~$\sim$~55~AU), \cite{GolimowskiD-04-AJ-2} (a~$\sim$~0.83--30~AU), \cite{Phan-Bao-05-A+A-3} (a~$\sim$~33~AU), \cite{Billeres-05-A+A-3} (a~$\sim$~220~AU), \cite{Law-06-MNRAS-4} (a~$\sim$~10--94~AU), \cite{CloseL-07-ApJ-4} (a~$\sim$~212--243~AU), \cite{ReidI-07-AJ-4} (a~$\sim$~100--3200~AU), \cite{KrausA-07-ApJ-4} (a~$\sim$~1600 AU), \cite{KonopackyQ-07-ApJ-2} (four out of 13 binaries have a~$>$~100~AU), \cite{RadiganJ-08-ApJ} (a~$\sim$~135~AU), \cite{RadiganJ-09-ApJ-2} (a~$\sim$~6700 AU), \cite{KrausA-09-ApJ-2} (a~$\sim$~500--5000~AU), \cite{BergforsC-10-A+A-2} (a~$\sim$~3--180~AU), \cite{LawN-10-ApJ-1} (a~$\sim$~600--6500~AU), \cite{JansonM-12-ApJ-3} (a~$\sim$~3--227~AU), \cite{DieterichS-12-AJ-2} (a~$\sim$~5--70~AU), \cite{JansonM-14-ApJ-4} (a~$\sim$~0.4--94~AU), \cite{DeaconN-14-ApJ-2} (a~$\sim$~300--69\,706~AU), \cite{Ward-DuongK-15-MNRAS-2} (a~$\sim$~3--10\,000~AU), \cite{Cortes-ContrerasM-17-A+A-2} (a~$\sim$~1--66 AU), \cite{Galvez-Ortiz-17-MNRAS} (a~$\sim$~200--92\,000~AU), \cite{Winters-19-AJ-4} (a~$\sim$~10$^{-2}$--10$^{4}$~AU). These discoveries of very wide separations of VLM binaries have been tried to explain recently with several models including the state of the art embrio-ejection scenario \citep[][]{Reipurth-01-AJ-4, Boss-01-ApJ-6}. Numerical simulations have produced results which have led to claims related to binary formation via ejection \citep[][]{BateM-03-MNRAS-3, UmbreitS-05-ApJ-1}.

Recently, \cite{Riaz-18-MNRAS-3} using the numerical simulations have attempted to model the initial stages of the formation of unequal mass brown dwarf and VLM systems evolving within a common envelope of gas. Starting from molecular cloud cores with various rotation rates and considering non-axisymmetric density perturbations to mimic the presence of large-scale turbulence on the level of molecular cloud cores, authors have successfully addressed the formation of wide VLM and brown dwarf binaries with semi-major axis up to 441 AU. It is to be noted that, since the binary separations for both of the binaries discussed in this paper (i.e. \lp1~AB and \lp8~AB) range below this limit, the formation mechanism of both of them could be well explained with this scheme. However, as discussed above, many of the VLM binaries discovered in recent years have a binary separation wider than 441 AU. Therefore those binaries could not be explained through the scheme of \cite{Riaz-18-MNRAS-3}. That might be due to the fact that the simulations of \cite{Riaz-18-MNRAS-3} did not take into account for radiative feedback effects on the collapsing gas, which have a profound impact on the star-forming gas \citep[][]{OffnerS-09-ApJ-2, BateM-12-MNRAS, MyersA-14-MNRAS-2, KrumholzM-16-MNRAS-2}. Further theoretical development is required in this context. 

In Figure~\ref{fig:sep_dist}, we have shown the separations of the VLM binaries \lp1\ and \lp8\ with red and blue solid vertical lines, respectively. We have also overplotted the separation-distribution of few latest works on VLM binaries. With the increase in the number of VLM binaries over the years, the peak of the separation distribution has also been changed. The studies of \cite{FischerD-92-ApJ-4}, \cite{Close-03-ApJ-4}, \cite{GizisJ-03-AJ-5}, and \cite{SieglerN-05-ApJ-2} indicated the peak of the distribution of separation of VLM binaries are $\sim$4 AU. For a sample of 0.1~\msun~$\lsimeq$~M~$\lsimeq$~0.5\msun, \cite{DucheneG-13-ARA+A-1} have derived a peak of separation distribution to be $\sim$5.3 AU. Recent studies of \cite{JansonM-14-ApJ-4} (see Figure~\ref{fig:sep_dist}; black dot-dot-dashed line) shows a distribution peak $\sim$6 AU, whereas \cite{Winters-19-AJ-4} with the confirmed 290 sample within 25 pc (see Figure~\ref{fig:sep_dist}; brown dotted line) have derived the peak of the separation distribution to be $\sim$20 AU. However, while taking a sub-sample of the stars lie within the 10~pc distance in solar neighbourhood (see Figure~\ref{fig:sep_dist}; violet dot-dashed line), authors have noticed that the distribution peak shifts to the value at $\sim$4 AU. In this analysis, the VLM binaries we have reported is within 1$\sigma_{log~a}$ of the total sample of the latest distribution of VLM binaries (see the green-shaded region in Figure~\ref{fig:sep_dist}). For comparison, in Figure~\ref{fig:sep_dist}, we have also plotted the separation distribution for solar-type stars \citep[peak $\sim$51 AU;][]{RaghavanD-10-ApJS-2}. Although the peak of solar-type stars also come within the 1$\sigma_{log~a}$, the separation of \lp8~AB is nearer to this value.

\subsection{Stability of the binary systems}
\label{subsec:disc_stability}

\begin{figure*}
\centering
\subfloat[]{\includegraphics[height=8.4cm, angle=-90]{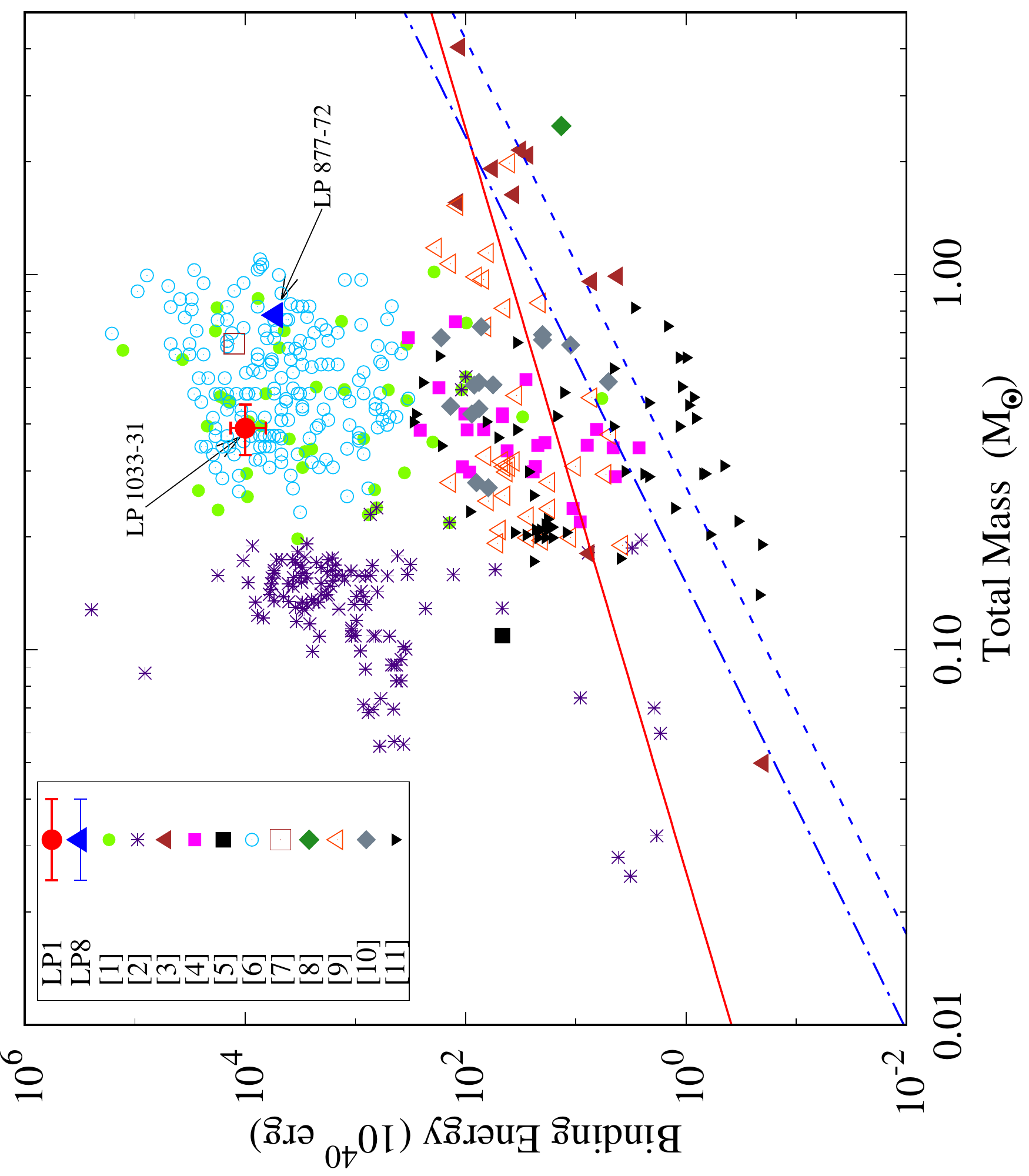}}
\hspace{0.2mm}
\subfloat[]{\includegraphics[height=8.4cm, angle=-90]{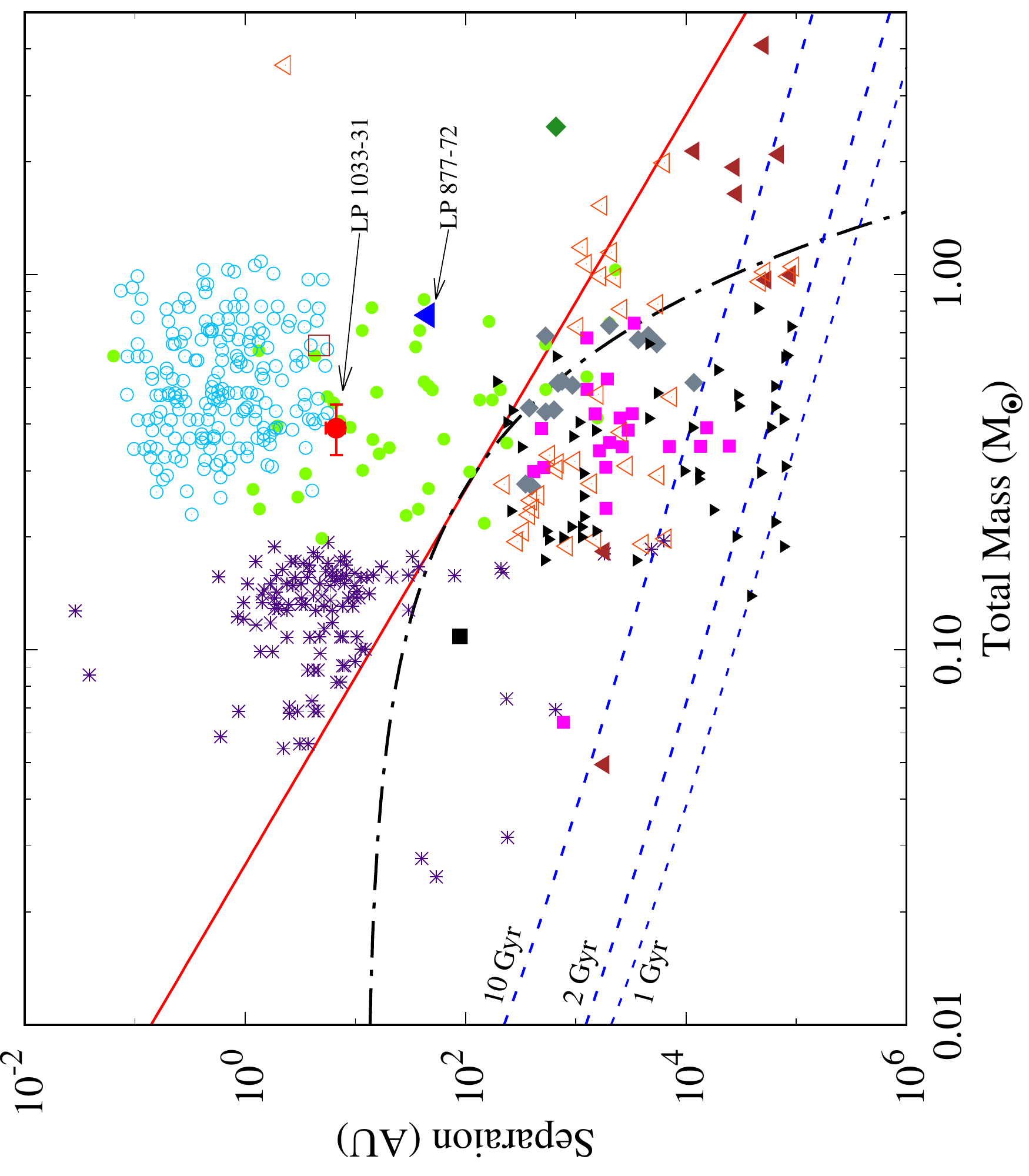}}
\caption{(a) System Binding Energy vs. Total Mass (primary + secondary) and (b) Binary Separation vs. Total Mass is shown. The detected binary systems \lp1~AB (shown as LP1) and \lp8~AB (shown as LP8) have been shown with red solid circle and blue solid triangle, whereas the known binary systems from the literature are marked with different symbols. The estimated uncertainties have also shown for both the systems. For \lp1~AB, the uncertainties are clearly visible, whereas for \lp8~AB the uncertainties are smaller than the size of the symbols. 
Figure legend: [1]: \pcite{FischerD-92-ApJ-4} and \pcite{ReidI-97-AJ-1}, [2]: VML archive, \pcite{BurgasserA-07-ApJ-5} and \pcite{SieglerN-05-ApJ-2}, [3]: \pcite{CaballeroJ-06-A+A-1} \& \pcite{Caballero-07-A+A-4, CaballeroJ-07-ApJ-2, CaballeroJ-09-A+A-1}, [4]: \pcite{LuhmanK-09-ApJ-7, LuhmanK-12-ARA+A-5}, \pcite{DhitalS-10-AJ-1} \& \pcite{MuzicK-12-AJ-1}, [5]: \pcite{BurninghamB-10-MNRAS-2}, [6]: \pcite{FahertyJ-10-AJ-3} \& \pcite{BaronF-15-ApJ-2}, [7]: Ross 458 AB, \pcite{BeuzitJ-04-A+A-1} and \pcite{GoldmanB-10-MNRAS}, [8]: HIP 78530 AB, \pcite{LafreniereD-11-ApJ}, [9]: \pcite{JansonM-12-ApJ-3}, [10]: \pcite{DeaconN-14-ApJ-2}, [11]: \pcite{Galvez-Ortiz-17-MNRAS}.
In the panel (a), we have overplotted the minimum binding energy line from \pcite{Zuckerman.-09-A+A-2} (red solid line) and the two lines showing the Jeans length criteria for mass ratio q = 1.0 (blue dot-dashed line) and q = 0.1 (blue dashed line) adopted from \pcite{FahertyJ-10-AJ-3}. 
In the panel (b), the empirical limit for stability stated by \pcite{ReidI-01-AJ-12} (black dot-dashed line) and by \pcite{BurgasserA-03-ApJ-8} (red solid line) have been shown. We also overplot the lifetime isochrones suggested by \pcite{DhitalS-10-AJ-1} with data of \pcite{CloseL-07-ApJ-4} and \pcite{WeinbergM-87-ApJ-2} equations for dissipation times of 1, 2 and 10 Gyr (blue dashed lines).
Both the systems \lp1~AB and \lp8~AB being well above the minimum-binding energy lines in (a) as well as the stability limits in (b) signifies that both systems are stable.
}
\label{fig:mass_sep_bind_energy}
\end{figure*}

Several studies have been carried out over the last few decades to investigate the stability of the binary systems in both theoretical and observational point of view. In Figure~\ref{fig:mass_sep_bind_energy}, we plot the diagram of total mass versus the binding energy (a) and total mass versus binary separation (b). In both the plot, same symbols have been used. The binaries \lp1\ and \lp8\ have been shown with a red solid circle and blue solid triangle. In the case of \lp8, the sizes of the error bars are smaller than the sizes of the symbols. We have also overplotted other M-dwarf binaries as well as VLM binaries in order to compare them with \lp1\ and \lp8. By close inspection, it is clear from both the figures that the values of \lp1\ and \lp8\ have more similarities to the studies by \cite{FischerD-92-ApJ-4}, \cite{ReidI-97-AJ-1}, \cite{BeuzitJ-04-A+A-1}, \cite{FahertyJ-10-AJ-3}, and \cite{BaronF-15-ApJ-2}. That would be due to the fact that both of these sources have similar spectral types and also have the same properties of the solar neighbourhood. 

\cite{Zuckerman.-09-A+A-2} derived a cutoff for the binding energy of a system formed by fragmentation as a function of the total system mass. In this calculation, authors assumed the binary separations of 300 AU. However, due to the findings of a number systems having separations larger than 300 AU, \cite{FahertyJ-10-AJ-3} used the Jeans length instead of the fiducial value of 300 AU to repeat the calculation for two extreme mass ratio cases. In Figure~\ref{fig:mass_sep_bind_energy}(a), we have shown the binding energy cutoff estimated by \cite{Zuckerman.-09-A+A-2} (red solid line) and \cite{FahertyJ-10-AJ-3} (blue dashed line for mass ratio = 0.1, and blue dot-dashed lines for Mass ratio = 1.0). Both \lp1\ and \lp8\ are found above the \cite{Zuckerman.-09-A+A-2} and \cite{FahertyJ-10-AJ-3} limits of bound systems.

Further, in Figure~\ref{fig:mass_sep_bind_energy}(b), we have also shown the suggested empirical limits for the stability of VLM multiples from the studies of \cite{ReidI-01-AJ-12} (black dashed line), and \cite{BurgasserA-03-ApJ-8} (red solid line), based on the objects that were known at the time of the respective works. However, none of these cutoffs seems to be appropriate for the collection of widely separated systems \citep[see][]{FahertyJ-10-AJ-3}. Later \cite{DhitalS-10-AJ-1} applied Galactic disc mass density \citep[][]{CloseL-07-ApJ-4} and other updated parameters to \cite{WeinbergM-87-ApJ-2} equations and described statistically the widest binary that is surviving at a given age. In the Figure~\ref{fig:mass_sep_bind_energy}(b), we have also overplotted the lifetime `isochrones' suggested by \cite{DhitalS-10-AJ-1} for the dissipation times of 1, 2 and 10 Gyr (blue dashed lines). With the parameters calculated in the previous sections, both systems \lp1\ and \lp8\ appear to be stable enough to survive longer than 10 Gyr. 

\subsection{Possibility of hosting exoplanets}
\label{subsec:disc_host_exoplanet}
In this section, we have discussed the possibility of hosting exoplanet on the identified VLM components using the latest investigations on similar type of stars. Utilising the M-dwarf surveys conducted with the HIRES/Keck, PFS/Magellan, HARPS/ESO, and UVES/VLT instruments supported with data from several other instruments \cite{TuomiM-19-arXiv} analysed the radial velocities of an approximate volume- and a brightness-limited sample of 426 nearby M-dwarfs and estimated that M-dwarfs (M0--M9.5) have on average at least 2.39$^{+4.58}_{-1.36}$ planets per star orbiting them. Using \textit{Kepler} data, \cite{Hardegree-Ullman-19-AJ-2} have also derived the planet occurrence rate specifically for mid-M-dwarfs (spectral types M3 V to M5.5 V) to be 1.19$_{-0.49}^{+0.70}$. \cite{Hardegree-Ullman-19-AJ-2} also estimated the planet occurrence rate for smaller ranges of spectral type M3 V -- M3.5 V, M4 V -- M4.5 V, and M5 V -- M5.5 V to be 0.86$^{+1.32}_{-0.68}$, 1.36$^{+2.30}_{-1.02}$, and 3.07$^{+5.49}_{-2.49}$ planets per star. 
Since the estimated values of planet occurrence rate by \cite{Hardegree-Ullman-19-AJ-2} show an increasing trend for larger spectral type, we might expect that for the spectral type $>$M5.5 the planet occurrence rate would be $\gsimeq$3.07 planets per star. Similarly, we also expect the planet occurrence rate for spectral types M1 and M2, to be lower than that was derived for spectral type M3 (i.e. the rate for planet occurrence would be $\lsimeq$0.86 planets per star).

In order to estimate the possibility of hosting planets for the detected VLM binaries, we first considered the planet occurrence rate for the individual components for each VLM binaries. After that we have also evaluated the effects of the binarity for each of the components. Considering the spectral types of \lp1~A ($\sim$M4.5), \lp1~B ($\sim$M4.5), \lp8~A ($\sim$M1), and \lp8~B ($\sim$M4), the planet occurrence rate would have been 0.34 -- 3.66, 0.34 -- 3.66, $\lsimeq$0.86, and 0.34 -- 3.66 planets per star. However, \cite{WangJ-14-ApJ-24} and \cite{KrausA-16-AJ-2} found that close binary companions appear to suppress planet formation and hence decrease planet occurrence rates for these systems. Recently, \cite{MoeM-19-arXiv-1} have compiled the recent works by \cite{WangJ-14-ApJ-24, WangJ-15-ApJ-35, WangJ-15-ApJ-5, KrausA-16-AJ-2, NgoH-16-ApJ-2, MatsonR-18-AJ-3, ZieglerC-20-AJ-2}, and have derived the relationship between the Suppression Factor (S$_{bin}$) as a function of binary separation. Using this relationship \citep[see Figure~3 of][]{MoeM-19-arXiv-1}, we have estimated the values of the S$_{bin}$ to be 0.126$\pm$0.013 and 0.5708$\pm$0.0018 for \lp1~AB and \lp8~AB, respectively. 
Therefore, considering the suppression due to the binarity, for each of the primary and secondary components of \lp1~AB, we derive that the probability of occurring planet would be the same and with the range of 4--51\%. In the case of \lp8, the primary and secondary components of the planet occurrence rate would be $\lsimeq$ 49\% and 19--208\%. This means that, if we search for exoplanets around each of the components of the VLM binaries reported in this paper, we should expect to find up to `two' exoplanets around \lp8~B, in contrast with the \lp8~A, \lp1~A, and \lp1~B, all three of the components have the maximum probability of hosting exoplanet are only $\sim$50\%.

\section{Summary and conclusions}
\label{sec:summary}
In this paper, we have presented a detailed study of two sources \lp1\ and \lp8, one of which was previously known as a single star (\lp1), whereas another source (\lp8) was recently resolved into two components by \textit{Gaia}. However, the properties of the newly discovered components were not well studied. Using the observations from NaCo/VLT high-resolution AO imaging in NIR $JHK_s$ band, we found that \lp1\ consists of two components  that have similar brightness, mass, radius, $\teff$, $\logg$, and luminosity (within its $\sim$1$\sigma$ uncertainties). However, in case of \lp8, these parameters are found to vary between the primary and secondary over a wide range. We have also investigated the properties of the binary systems, and the results are summarised below:
\begin{itemize}
\item The PA and the projected physical separation of \lp1~AB are derived to be 334.01\deg$\pm$0.12\deg\ and 6.7$\pm$1.3 AU. The same parameters for \lp8\ are estimated to be 138.60\deg$\pm$0.07\deg\ and 45.8$\pm$0.3~AU, respectively. 

\item The spectral types for both the primary and secondary of \lp1~AB are found to be similar and estimated to be M4.5. Whereas for \lp8~AB, the spectral types of the primary and the secondary components are derived to be $\sim$M1 and $\sim$M4. 
  
\item The system \lp1~AB is found to have two nearly equal mass components with the estimated masses of 0.20$\pm$0.04 and 0.19$\pm$0.04~\msun. Whereas for \lp8~AB, the masses are estimated to be 0.520$\pm$006 and 0.260$\pm$0.006~\msun.

\item The primary and the secondary components of \lp1~AB are found to have similar radius and luminosity, with the values of $\sim$0.22~\rsun\ and $\sim$0.005~\lsun, respectively. Whereas for \lp8, the radius of the primary is twice the radius of the secondary with the values of $\sim$0.49 and $\sim$0.27~\rsun. However, the primary is five times more luminous than the secondary of \lp8~AB with the lumionosity of the primary to be $\sim$0.043~\lsun.

\item For \lp1, the $\teff$ and $\logg$ are found to be similar with the values of $\sim$3200~K and $\sim$5.06. In the case of \lp8, the $\teff$ of the primary is $\sim$3750~K, which is $\sim$400~K hotter than the secondary component. The $\logg$ of the secondary component is found to be 1.05 times more than the primary component with a log~$g$ of the primary being 4.768.

\item The orbital period of \lp1\ and \lp8\ are also estimated to be 28$\pm$3 and 349$\pm$3 yr, respectively. We have estimated the binding energy for \lp1\ and \lp8\ to be --1.00$\pm$0.35~$\times$~10$^{44}$ and --5.21$\pm$0.12~$\times$~10$^{43}$ erg, respectively. We have also shown that both the systems are stable enough to survive longer than $\sim$10 Gyr.

\item We have discussed the possibility of hosting exoplanets on each of the components of the VLM binaries. We should expect to find up to `two' exoplanets around \lp8~B, whereas the maximum probability for hosting exoplanets for \lp8~A, \lp1~A, and \lp1~B are estimated to be $\sim$50\%. 

\end{itemize}
In future, the follow-up spectroscopic observations of the detected individual components for both \lp1\ and \lp8\ can be made with AO. This would be very useful for further understanding of the onset of dust cloud formation in their atmosphere. The detection and characterisation of the two VLM binaries including an equal mass binary, give us the opportunity to constrain the evolutionary models as well as to perform the follow-up radial velocity observations for the possible detection of the exoplanets around them.
\section*{Acknowledgments}
We thank the anonymous referee for a rigorous review that greatly improved the manuscript.
SK and AR are especially grateful to Sachindra Naik and Mudit Kumar Srivastava from \textit{Physical Research Laboratory} (PRL) for their valuable inputs and discussion.
The research leading to these results has received funding from the French ``Programme National de Physique Stellaire'' and the Programme National de Planetologie of CNRS (INSU). The computations were performed at the {\sl P\^ole Scientifique de Mod\'elisation Num\'erique} (PSMN) at the {\sl \'Ecole Normale Sup\'erieure} (ENS) in Lyon, and at the {\sl Gesellschaft f{\"u}r Wissenschaftliche Datenverarbeitung G{\"o}ttingen} in collaboration with the Institut f{\"u}r Astrophysik G{\"o}ttingen. DH was supported by the Collaborative Research Centre SFB 881 ``The Milky Way System'' (subproject A4) of the German Research Foundation (DFG).

\section*{Data availability}
The data underlying this article are available in the ESO archive (\href{https://archive.eso.org/eso/eso_archive_main.html}{https://archive.eso.org/eso/eso\_archive\_main.html}).

\bibliographystyle{mnras}
\bibliography{ref,SK_collections}
\end{document}